\documentclass[aps,prx,twocolumn,showpacs,showkeys,nofootinbib,tightenlines,nobibnotes,superscriptaddress]{revtex4-1}
\usepackage{graphicx}     
\usepackage{amsmath}
\usepackage{amssymb, amsfonts}
\usepackage{array}
\newcolumntype{x}[1]{>{\centering\let\newline\\\arraybackslash\hspace{0pt}}p{#1}}
\usepackage{braket}
\usepackage{cases}
\usepackage{bm}
\usepackage[utf8]{inputenc}
\usepackage[overload]{empheq}
\usepackage{relsize}  
\usepackage{multirow}
\usepackage{dcolumn}  
\usepackage{hyperref}
\usepackage{xcolor}
\usepackage{relsize}%
\hypersetup{
	colorlinks   = true, 
	urlcolor     = blue, 
	linkcolor    = blue, 
	citecolor    = blue 
}
\newcommand{\RN}[1]{%
	\textup{\uppercase\expandafter{\romannumeral#1}}%
}
\newcommand{\angstrom}{\text{\normalfont\AA}}
\newcolumntype{d}[1]{D{.}{.}{#1}}
\newcolumntype{K}[1]{>{\centering\arraybackslash}p{#1}}

\usepackage{floatflt}     
\usepackage{wrapfig}
\usepackage{color}
\usepackage{pstricks}

\newcommand{\overbar}[1]{\mkern1.5mu\overline{\mkern-1.5mu#1\mkern-1.5mu}\mkern 1.5mu}
\makeatletter
\newcommand*{\rom}[1]{\expandafter\@slowromancap\romannumeral #1@}
\graphicspath{{./eps/}{./png/}}
\usepackage{color}
\definecolor{DarkRed}{rgb}{0.35,0.01,0.01}
\definecolor{Linen}{rgb}{0.98,0.98,0.94}
\definecolor{Blue}{rgb}{0.,0.,1.0}
\definecolor{DarkBlue}{rgb}{0.099,0.099,0.44}
\definecolor{DarkGreen}{rgb}{0.0,0.4,0.0}
\definecolor{Turquoise}{rgb}{0.0,0.9,0.7}
\begin{document}
\title{Pico-photonics: Anomalous Atomistic Waves in Silicon}
\author{Sathwik Bharadwaj}
\affiliation{Birck Nanotechnology Center, School of Electrical and Computer Engineering, Purdue University, West Lafayette, Indiana 47907, USA}
\author{Todd Van Mechelen}
\affiliation{Birck Nanotechnology Center, School of Electrical and Computer Engineering, Purdue University, West Lafayette, Indiana 47907, USA}
\author{Zubin Jacob}
\email{zjacob@purdue.edu}
\affiliation{Birck Nanotechnology Center, School of Electrical and Computer Engineering, Purdue University, West Lafayette, Indiana 47907, USA}
\begin{abstract} 
The concept of photonic frequency $(\omega)$ - momentum $(q)$ dispersion has been extensively studied in artificial dielectric structures such as photonic crystals and metamaterials. However, the $\omega-q$ dispersion of electrodynamic excitations hosted in natural materials at the atomistic level is far less explored. Here, we develop a Maxwell Hamiltonian theory of matter combined with
the quantum theory of atomistic polarization to obtain the electrodynamic dispersion of natural materials interacting with the photon field. We apply this theory to silicon and discover the existence of anomalous  atomistic waves. These waves occur in the spectral region where propagating waves are conventionally forbidden in a macroscopic theory.  Our findings demonstrate that natural media can host a variety of yet to be discovered waves with sub-nano-meter effective wavelengths in the pico-photonics regime.

\end{abstract}
\maketitle
\begin{center}
\today
\end{center}
\section{Introduction}\label{sec:intro}
Functional dependency of the energy and momentum (dispersion) of particles hosted in matter captures the fundamental properties of a material \cite{cohen_louie_2016}.  The dispersion for several  electronic, phononic, and magnonic excitations in condensed matter systems \cite{yu2010fundamentals, misra2011physics, Togo_phononbandstructure, Fransson_magnontopology, Vasseur_magnonbandstructure, Chisnell_magnontopology,Barbara_fullerene_exciton, Fengcheng_excitonMos2, Cheiwchanchamnangij, Chi-Cheng_exciton, Farid_plasmondispersion, Fabio_Polaron, Stephan_polaron, Bulut_quasiparticlesuperconductorband, Norman_superconductorband} have  been widely studied within an atomistic lattice band theory. However, the concept of frequency and momentum $(\omega-q)$ photonic dispersion \cite{johnson_joannopoulos_2005} and the corresponding electromagnetic field confinement \cite{Hugonin} have been formulated only in artificial materials such as photonic crystals \cite{Robertson_pcmeasurement, Yablonovitch_pcfcc, Siying_topologicalpc, Ozawa_topological_photonics}, metamaterials \cite{ARaman_metbandstructure, Orlov_dispersionMeta, Coevorden_dipolarlattice}, and other dielectric structures \cite{Soukoulis_pcgapperiodic, Yablonovitch_nonsphericalpc}. These artificial materials are composed of two or more macroscopic constituents. On the contrary, natural media itself can host  electrodynamic excitations which adapt the symmetry and periodicity of the material \cite{Todd_nonlocal_topology_em, Todd_sathwik_topologyobstruction}. Hence, natural materials can host a variety of yet to be discovered electrodynamic waves and topological photonic properties \cite{Todd_viscous_maxwell, mechelen_sun_jacob_2021, Todd_sathwik_topologyobstruction}. As such, these are the properties of matter itself and are not related to a form of macroscopic engineering.  In this article, we develop a Maxwell Hamiltonian theory of matter combined with the quantum theory of atomistic polarization to unveil the electrodynamic dispersion of the electromagnetic (photon) field.   
\begin{figure*}[t]
    \centering
    \includegraphics[width = 6in]{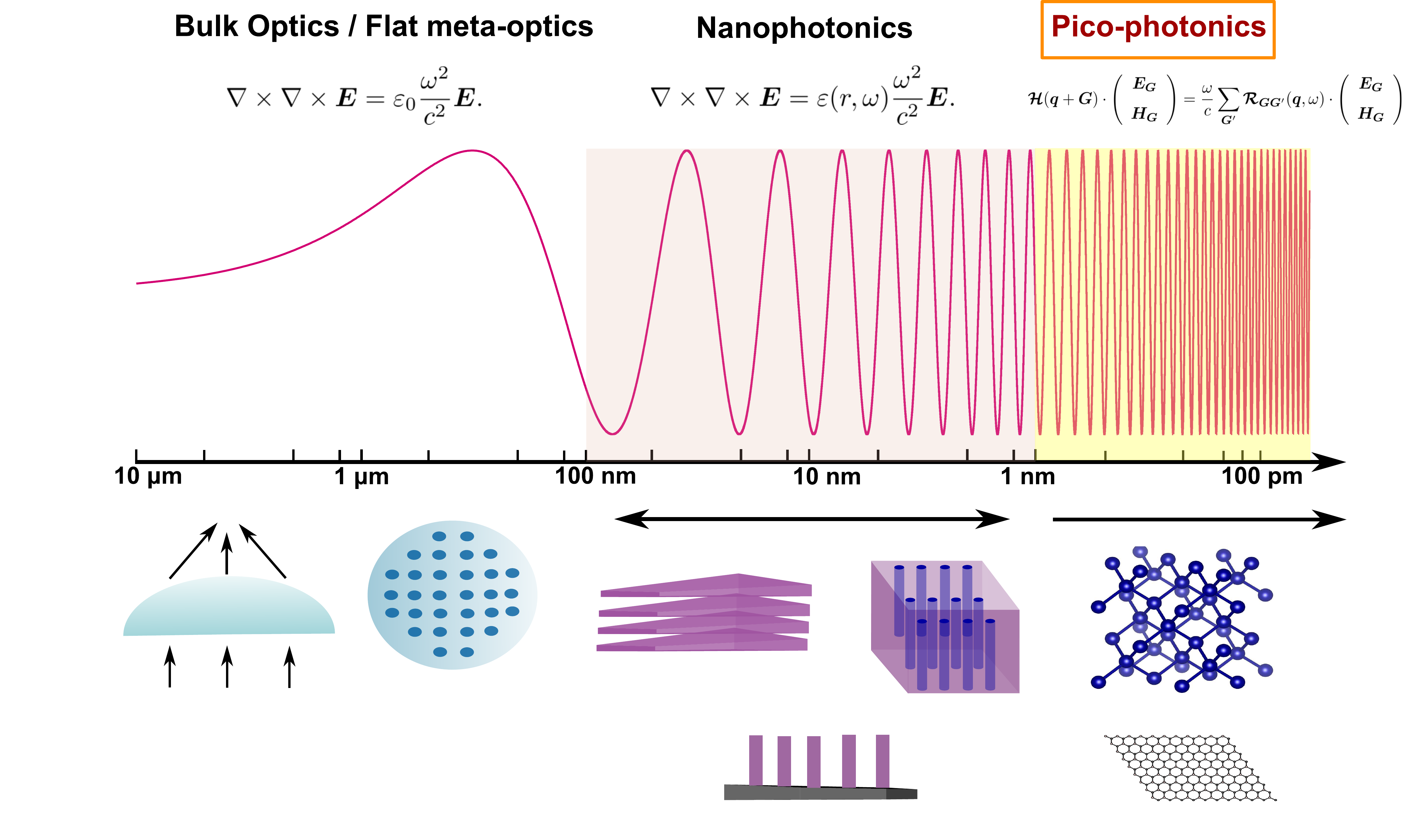}
    \caption{Branches of optics across the length-scales are depicted in the schematic diagram. Bulk optics and flat meta-optics are applicable for electromangetic waves passing through micro-meter scale artificial structures. Light-matter interaction in metamaterials, photonic crystals are studied within nanophotonics. Here, we define the field of pico-photonics, where we analyze the light-matter interaction in natural materials at sub-nm regime. Field equations in bulk optics satisfy a standard Maxwell wave equation with a constant dielectric permittivity. In nanophotonics, the dielectric function may depend on frequency and vary spatially over the region of interest. Field solutions in nanophotonics regime still satisfy the classical wave equation. On contrary, in pico-photonics, the response function is dependent on frequency, momentum, and the local-field effects, and the dynamics of electromagnetic waves are studied within the Maxwell Hamiltonian framework.}
    \label{fig:perceptive}
\end{figure*}

Recently, it has been shown that a graphene monolayer in the viscous hydrodynamic state \cite{MPolini_nonlocalHall} supports spin-1 skyrmions in the bulk and topologically protected electromagnetic edge states at the boundary \cite{mechelen_sun_jacob_2021, VanMechelenJacob_Maxwell, VanMechelenOpticsexpress}. This topological electrodynamic phase of matter is characterized by an optical $N$-invariant \cite{Todd_sathwik_topologyobstruction} fundamentally distinct from the Chern number and $\mathbb{Z}_2$ invariant. The optical N-invariant was defined based on a semi-classical hydrodynamic nonlocal (photon momentum $\hbar\bm{q}\neq 0$) dielectric response which includes Hall viscosity and the dynamics of electromagnetic waves. However, the Maxwell Hamiltonian theory of matter and the quantum theory of atomistic polarization within the framework of a lattice band theory has not been considered so far. Here, we solve this key challenge and show that the atomistic polarization results in the emergence of unique class of atomistic waves. In this paper, we apply the theory to silicon but in the future it can be adopted to topological systems with repulsive Hall viscosity.

In Fig.~\ref{fig:perceptive}, we compare the light-matter interaction theories across varying length scales. Tradition regime of optics and flat meta-optics study the optical properties within the classical electromagnetism, and the dielectric response is considered to be a material dependent function. Nanophotonics encompass the study of electromagnetic field interactions in artificial structures such as metamaterials, photonic crystals, and other dielectric structures \cite{mcgurn_2019}. Field solutions in these structures can be effectively obtained through a classical wave equation, with the dielectric response dependent on the spatial geometry \cite{PandeySathwik_EM} and frequency. In this article, our focus is pico-photonics, which comprises the light-matter interaction in natural materials at sub-nano-meter (nm) regime. We show that in the pico-photonic regime, the electromagnetic fields satisfy a pico-photonic Bloch functional form. Dynamics of the fields are defined by a pico-photonic nonlinear eigenvalue equation, which depends on the quantum theory of atomistic polarization as opposed to semi-classical Drude or hydrodynamic models. Further, we apply this formulation for Si, and discover the existence of anomalous atomistic waves. These waves occur in the frequency range where propagating waves are conventionally forbidden in a macroscopic theory. We show that the anomalous waves observed in Si are highly oscillatory within a unit cell, well within the dominion of pico-photonics.  

The paper is arranged as follows. In Sec.~\ref{sec:atomisticdielectric}, we define the atomistic dielectric tensor and and discuss the importance of contributions from the local-field effects in a material.  In Sec.~\ref{sec:microscopicdielectric}, we derive the transverse atomistic dielectric tensor within a linear response theory. An atomistic nonlocal electrodynamics of matter based on the Maxwell Hamiltonian is described in Sec.~\ref{sec:atomisticEMtheory}. In this section, we also define the pico-photonic bloch function and the pico-photonic eigenvalue equation for the electrodynamic field. As an application of our formulation, we obtain the nonlocal atomistic dielectric response and the corresponding atomistic electrodynamic dispersion in Si through an isotropic nearly-free electron model, as described in Sec.~\ref{sec:isotropicmodel} and Sec.~\ref{sec:atomisticEMbandstructure}, respectively. Concluding remarks are presented in Sec.~\ref{sec:conclusions}. 

\section{Defining the atomistic dielectric tensor}\label{sec:atomisticdielectric}
In solid-state materials, long-wavelength perturbations can lead to short-wavelength responses due to short range electronic correlations \cite{Sinha_localfield, Hybertsen_static_dielectric, CohenLouieChelikowsky}. This phenomenon has been termed as the local-field effect \cite{HankeSham}. Consequently, microscopic fields arising from the local-field effects vary rapidly within the unit-cell. The macroscopic field is obtained through averaging the microscopic fields over a region large compared to the lattice constant. This macroscopic field is not the same as the atomistic electromagnetic field in a material \cite{Adler}. Inside a material, fields will have rapidly varying terms with wavevector $\bm{q}+\bm{G}$, where $\bm{G}$ is the reciprocal lattice vector and $\bm{q}$ is the photon wavevector. Hence, the dielectric response of a material depends on frequency $(\omega)$, momentum $(\hbar \bm{q})$ and the local-field effects. The dielectric response of a material is represented in momentum space as
\begin{equation}
\varepsilon(\bm{q}+\bm{G}, \bm{q}+\bm{G}', \omega) \equiv \varepsilon^{\bm{G}\bm{G}'}(\bm{q}, \omega).
\end{equation}
When $\bm{q} \neq 0$, we obtain the nonlocal dielectric response, and the components with $\bm{G}, \bm{G}' \neq 0$ are due to local-field effects. So far, in literature, only the longitudinal dielectric function (density-density response) \mbox{$\varepsilon_L^{\bm{G}\bm{G}'}(\bm{q}, \omega)$} has been extensively studied \cite{Adler, NWiser}. However, a crucial gap in the linear response theory of matter is in understanding the influence of local-field effects on the dielectric response arising from a photon field. 

Traditionally, electromagnetic properties of matter are treated within a macroscopic local electrodynamic framework, where it is assumed that the dielectric function is only dependent on frequency $\varepsilon(\omega)$. This consideration is valid only in the long-wavelength limit, $\bm{q} \rightarrow 0$. Although there have been efforts to develop quantum-electrodynamic first-principles density-functional theory calculations \cite{Rubio_QED_cavity, Rubio_QEDFT}, applications of such frameworks have been limited to artificial dielectric structures and cavities. These frameworks are also developed in the long-wavelength limit and the photon field is considered to be in vacuum. In this article, our focus is the pico-photonic, atomistic regime beyond the cavity quantum electrodynamics and local dielectric response approximations.  

For a system with infinitesimal translation symmetry, a jellium model can be used, where we consider a nonlocal dielectric function $\varepsilon(\bm{q}, \omega)$ without any contribution from the local-field off-diagonal components. This approximation has been successfully applied for the case of simple metals \cite{Economou2010}. The jellium model breaks down in explaining the observed properties of nanoplasmonic structures with metals in the sub-nm domain \cite{Rubio_quantum_plasmonics}.  Nonlocal quantum effects in nanoplasmonic structures can be explained through hydrodynamic models \cite{Asger_nonlocalhydro, KamandarDezfouli:17} as opposed to a local Drude model response theory. However, as shown in this article, in semiconducting materials, the local-field effects beyond the hydrodynamic model takes the central role in determining the atomistic electrodynamic dispersion of matter. 

Early efforts within classical electrodynamics to include the local-field effects in the dielectric function were considered through the Clausius–Mossotti relation (Lorentz–Lorenz equation) \cite{frohlich_1990, Hannay_1983, Rysselberghe}. In this approximation, the simple cubic lattice of polarizable atomic sites is replaced with a homogeneous cavity. This leads to a connection between the macroscopic dielectric function $\varepsilon_M$ in terms of the molecular polarizability \cite{Aspnes}. However, the Clausius–Mossotti relation neither has frequency or momentum dependency of the dielectric function, and does not build in the symmetry of the Brillouin zone of the system. We also note that the widely used approximation of replacing atoms by polarizable harmonic oscillators is confined to the classical regime. Adler \cite{Adler} and Wiser \cite{NWiser} (from now on termed as the Adler-Wiser formulation) put forth the quantum theory of atomistic longitudinal dielectric function \mbox{$\varepsilon_L^{\bm{G}\bm{G}'}(\bm{q}, \omega)$} based on the perturbation theory. Following these efforts, it has been shown that the local-field corrections to \mbox{$\varepsilon_L^{\bm{G}\bm{G}'}(\bm{q}, \omega)$} are quintessential to determine the electron self-energy \cite{Govoni_WEST, Seitz_selfenergy, HybertsenGW} and impurity screening potential \cite{Selloni, Chelikowsky_exitonbinding}. Here, we introduce the transverse dielectric function $\varepsilon_T^{\bm{G}\bm{G}'}(\bm{q}, \omega)$ going beyond the Adler-Wiser formulation. 
\section{Beyond Adler-Wiser Formulation: Atomistic Dielectric Response in Matter}\label{sec:microscopicdielectric}
The Adler-Wiser formulation determines the atomistic longitudinal dielectric response including the local-field effects [ref]. This expression for the dielectric function has been the gold standard in first-principles calculations to determine the optical response of a material \cite{Bechstedt, QE_simplecode}. However, response of a material to a photon field is determined by the atomistic transverse dielectric tensor. In this section, we develop a quantum theory of $\varepsilon_T^{\bm{G}\bm{G}'}(\bm{q}, \omega)$, including the local-field effects. 

The dielectric function of a material can be expressed in a longitudinal and transverse basis \cite{Adler, Rajagopal, RajagopalJain} as
\begin{equation}
\varepsilon^{\bm{G}\bm{G}'}(\bm{q}, \omega) = \left[\begin{array}{cc}
\varepsilon_L^{\bm{G}\bm{G}'}(\bm{q}, \omega)     &  \varepsilon_{LT}^{\bm{G}\bm{G}'}(\bm{q}, \omega) \\[6pt]
\varepsilon_{TL}^{\bm{G}\bm{G}'}(\bm{q}, \omega)     & \varepsilon_{T}^{\bm{G}\bm{G}'}(\bm{q}, \omega)
\end{array}\right],
\end{equation}
where, $\varepsilon_L^{\bm{G}\bm{G}'}$ is the longitudinal dielectric response (density-density correlation), $\varepsilon_T^{\bm{G}\bm{G}'}$ is the transverse dielectric function (current-current correlation). The cross-coupling terms $\varepsilon_{LT}^{\bm{G}\bm{G}'}$ and $\varepsilon_{TL}^{\bm{G}\bm{G}'}$ represent the longitudinal and transverse dielectric response induced by the transverse and longitudinal field, respectively. However, in a cubic material such as Si, contributions from $\varepsilon_{TL}^{\bm{G}\bm{G}'}$ and $\varepsilon_{LT}^{\bm{G}\bm{G}'}$ are negligibly small \cite{DelSole, NWiser}, and are neglected from consideration.  

In Fourier space, the induced potential $\delta V_{\rm ind}(\bm{r},t)$ in a material due to an external potential $\delta V_{\rm ext}\left(\bm{r},t\right)$ can be expressed in terms of the longitudinal dielectric function  $\varepsilon_L^{\bm{G}\bm{G}'}$ as
\begin{equation}\label{eq:vextvrelation}
\delta V_{\rm ext}(\bm{q}+\bm{G}, \omega) = \sum_{\bm{G}'}\varepsilon_L^{\bm{G}\bm{G}'}(\bm{q}, \omega)\,\delta V_{\rm ind}(\bm{q}+\bm{G}', \omega), 
\end{equation} 
Whereas, $\varepsilon_T^{\bm{G}\bm{G}'}$ is defined as 
\begin{align}
\frac{4\pi c}{\omega^2}&\bm{J}_{\rm ind}(\bm{q}+\bm{G}, \omega) \nonumber\\&= \sum_{\bm{G}'}\left[\varepsilon_T^{\bm{G}\bm{G}'}(\bm{q}, \omega)-\delta_{\bm{G}\bm{G}'}\right]\cdot \bm{A}(\bm{q}+\bm{G}', \omega), 
\end{align}
where, $\bm{J}_{\rm ind}$ is the induced current and $\bm{A}$ is the transverse vector potential. We note that the transverse part of the vector potential $\bm{A}$ is gauge invariant. 
\subsection{Adler-Wiser Longitudinal Dielectric Function}
In literature, the longitudinal dielectric function is extensively studied including the local-field effects. We neglect the exchange-correlation contribution within the relaxation time approximation (RPA) \cite{Hybertsen_static_dielectric}. Our main contribution in this section is the transverse atomistic dielectric function. However, for completeness, we re-state the longitudinal dielectric function which is given by (see supplementary information for detailed derivation)
\begin{widetext}
\begin{align}
\varepsilon_L^{\bm{G}\bm{G}'}&(\bm{q},\omega) 
=\nonumber\\ &\delta_{\bm{G}\bm{G}'}-\frac{4\pi e^2}{q^2}\frac{1}{\Omega}\sum_{n,n',\bm{k}\sigma} f_{n\bm{k}} \left(1-f_{n'\bm{k+q}}\right) \left[\frac{\left<n,\bm{k}\right|e^{-i\left(\bm{q}+\bm{G}\right)\cdot\bm{r}}\left|n',\bm{k}+\bm{q}\right>\left<n',\bm{k}+\bm{q}\right|e^{i\left(\bm{q}+\bm{G}'\right)\cdot\bm{r}'}\left|n,\bm{k}\right>}{\left(\epsilon_{n,\bm{k}}-\epsilon_{n',\bm{k}+\bm{q}}+\hbar\omega+i\hbar\alpha\right)} + c.c\right],
\end{align}
\end{widetext}
where, $\Omega$ is the crystal volume, $\bm{k}$ and $\sigma$ are the carrier momentum and spin, $f_{n\bm{k}}$ is the Fermi-Dirac distribution, $n, n'$ are the band indices, and $\epsilon_{nk}$ is the eigen-energy. Conservation of crystal momentum has been built in the expression for the dielectric function. 

This response determines the plasmon screening in a material. Also, the screened coulomb interaction and the self-energy operator are determined by the above nonlocal longitudinal dielectric response function \cite{Govoni_WEST}. Hence, in $GW$ calculations, $\varepsilon_L^{\bm{G}\bm{G}'}(\bm{q}, \omega)$ are determined including the local-field effects. In Sec.~\ref{sec:atomisticEMtheory}, we employ $\varepsilon_L^{\bm{G}\bm{G}'}(\bm{q}, \omega)$ to determine the atomistic plasmon dispersion.  

\subsection{Beyond Longitudinal Dielectric Function: Transverse Dielectric Response}
We emphasize that the atomistic transverse dielectric function has received far less attention in literature. The behavior of propagating electrodynamic waves (i.e photons) is governed by the transverse response of matter.  Previous work from Adler derived the transverse dielectric function by assuming both the field and induced current density as macroscopic quantities \cite{Adler, Wilkins_quartz}. Here, we include all atomistic local-field contributions of the vector field and obtain the transverse dielectric function starting from the fundamental light-matter interaction Hamiltonian
\begin{equation}
H = \frac{\left(\bm{p}-\displaystyle\frac{e}{c}\bm{A}\right)^2}{2m} + U(\bm{r}),
\end{equation}
where $U(\bm{r})$ is the periodic lattice potential. Both $\bm{J}_{\rm ind}$ and $\bm{A}$ are microscopic in nature with components varying rapidly within the unit cell. Hence, the vector potential is of the form
\begin{equation}
\bm{A}(\bm{r}', \omega) = \sum_{\bm{G}',\bm{q}} A_{\bm{G}'}(\bm{q}, \omega)\,\bm{t}_{\bm{G}'}\, e^{i\left(\bm{q}+\bm{G}'\right)\cdot\bm{r}'},
\end{equation}
where, $\bm{t}_{\bm{G}}$ is the unit vector component perpendicular to $\bm{q}+\bm{G}$. 
In the supplementary information, we have derived $\varepsilon_T^{\bm{G}\bm{G}'}(\bm{q},\omega)$. Here, we state the important contribution of our manuscript which is 
\begin{widetext}
\hspace{-1.in}
\begin{align}
\varepsilon_T^{\bm{G}\bm{G}'}(\bm{q},\omega) = \delta_{\bm{G}\bm{G}'}+ \frac{4\pi e^2}{\Omega \omega^2}\sum_{n,n',\bm{k}}&\left<{n\bm{k}}\right|e^{-i\left(\bm{G}+\bm{q}\right)\cdot\bm{r}}\bm{t}_{\bm{G}}\cdot\bm{J}_0\left|{n'\bm{k}+\bm{q}}\right>\left<{n'\bm{k}+\bm{q}}\right|e^{i\left(\bm{G}'+\bm{q}\right)\cdot\bm{r}'}\bm{t}_{\bm{G}'}\cdot\bm{J}_0\left|{n\bm{k}}\right>\!\times\!\nonumber\\&\!{\left(f_{n'\bm{k}+\bm{q}}-f_{n\bm{k}}\right)}\!\left[{\rm P. V.}\left(\frac{1}{{\epsilon_{n'\bm{k}+\bm{q}}-\epsilon_{n\bm{k}}-\hbar\omega}}\right)+i\pi \delta\left({\epsilon_{n'\bm{k}+\bm{q}}-\epsilon_{n\bm{k}}-\hbar\omega}\right)\right],
\end{align}
\end{widetext}
where, $\bm{J}_0$ is the probability current operator. 
In Sec.~\ref{sec:atomisticEMtheory}, we show that this atomistic transverse dielectric response determines the pico-photonic dispersion of a material.
In Sec.~\ref{sec:isotropicmodel}, we apply these formulae to obtain the longitudinal and transverse dielectric function of Si based on an isotropic nearly-free electron bandstructure. 
\begin{figure*}
    \centering
    \includegraphics[width = 5in]{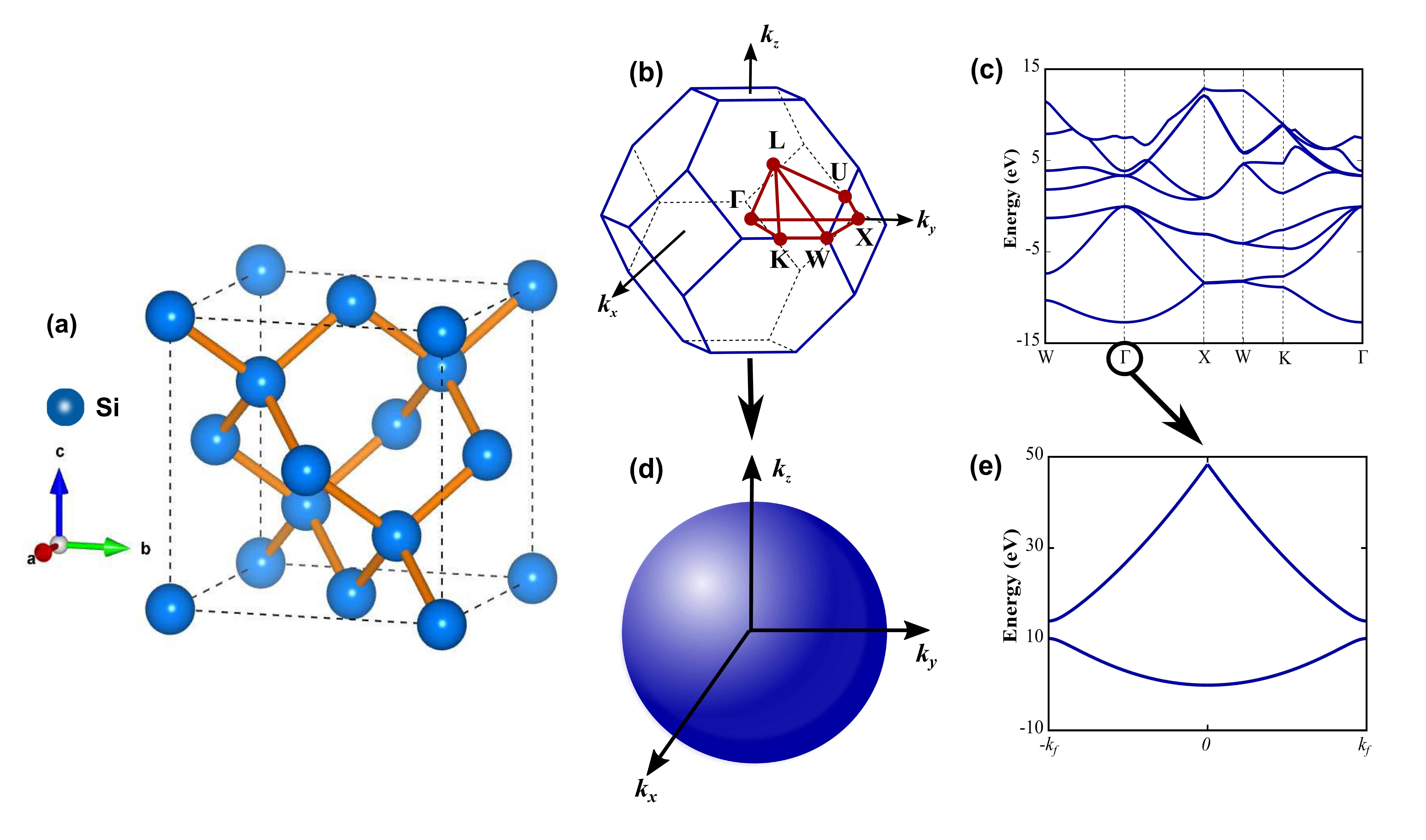}
    \caption{(a) Cubic crystal structure of silicon is shown. (b) First Brillouin zone of silicon, a truncated octahedron is plotted in k-space. (c) Spherical Brillouin zone used in this work to obtain the dielectric properties is plotted in k-space. (d) Bandstructure of silicon obtained using the empirical pseudo-potential method is displayed. (e) Bandstructure of silicon within a nearly-free electron model is displayed. This isotropic model can be thought of as a symmetric expansion of the bandstructure around the high-symmetric $\Gamma$ point.}
    \label{fig:bz}
\end{figure*}

\section{Maxwell Hamiltonian in Matter}\label{sec:atomisticEMtheory}
In this section, we develop the atomistic nonlocal electrodynamic theory of matter. We derive the Maxwell Hamiltonian in matter which depends on the spin-1 behavior of photons, analogous to the Dirac Hamiltonian for spin-1/2 particles. This formalism will be employed in the next section to obtain the  atomistic electrodynamic $\omega-q$ dispersion of a material. We emphasize that the Maxwell Hamiltonian has been used to understand the correspondence between photons and massless fermions in the Dirac equation specifically in free space. Only recently, the Maxwell Hamiltonian has regained attention in condensed matter to predict new topological electrodynamic phases of matter \cite{Todd_nonlocal_topology_em}. Our goal is to develop the Maxwell Hamiltonian formalism and apply it to a semiconducting material: silicon for the first time.

\subsection{Pico-photonic Bloch Function}
Atomistic electrodynamic dispersion of matter is obtained through solutions to the Maxwell Hamiltonian corresponding to  the  transverse  part  of  the  electromagnetic fields. The equation of motion for the Maxwell Hamiltonian $\bm{\mathcal{H}}$ (in Gaussian units) in vacuum (see Appendix~\ref{appdx:MaxwellHamiltonian}) is given by
\begin{align}\label{eq:EMhamiltonian}
\bm{\mathcal{H}}\cdot\bm{f} &= \frac{\omega}{c}\bm{g};\nonumber\\ \bm{f} = \left[\begin{array}{c}
     \bm{E}_T(\bm{r},\omega) \\[3pt]
      \bm{H}_T(\bm{r},\omega)
\end{array}\right],\,&\, \bm{g}= \left[\begin{array}{c}
     \bm{D}_T(\bm{r},\omega) \\[3pt]
      \bm{B}_T(\bm{r},\omega)
\end{array}\right],
\end{align}
where, 
\begin{equation}
\bm{\mathcal{H}} = \left[\begin{array}{cc}
0     &  \mathcal{H}^{\dagger}\\
\mathcal{H}     & 0
\end{array}\right];\quad \mathcal{H} = \bm{q}\cdot\bm{\mathcal{S}}. 
\end{equation}
Here, $\bm{q} = -i\bm{\nabla}$ is the momentum operator, $\bm{\mathcal{S}}$ is the spin-1 operator. The Maxwell Hamiltonian is expressed in terms of spin-1 operators of photon \cite{berry_spin1} and the components of the spin-1 operators are defined as
\begin{align}
\mathcal{S}_x = \left[\begin{array}{ccc}
0 & 0    & 0 \\
0 &   0  &  -1\\
0 & 1    & 0
\end{array}\right];\,\,&\,\, \mathcal{S}_y = \left[\begin{array}{ccc}
0 & 0    & 1 \\
0 &   0  &  0\\
-1 & 0    & 0
\end{array}\right];\nonumber\\&\\ \mathcal{S}_z =& \left[\begin{array}{ccc}
0 & -1    & 0 \\
1 &   0  &  0\\
0 & 0    & 0
\end{array}\right], \nonumber     
\end{align}
and they satisfy the angular momentum algebra $\left[\mathcal{S}_i, \mathcal{S}_j\right] = \epsilon_{ijk}\mathcal{S}_k$. Given a translation operator $\mathcal{T}$, the field vector $\bm{f}(\bm{r}, \omega)$ and the displacement vector $\bm{g}(\bm{r}, \omega)$ follow the relation
\begin{align}
\mathcal{T}\cdot\bm{f}(\bm{r}, \omega) &=  \bm{f}(\bm{r}+\bm{R}, \omega),\nonumber\\ \mathcal{T}\cdot\bm{g}(\bm{r}, \omega) &=  \bm{g}(\bm{r}+\bm{R}, \omega).   
\end{align}
It is easy to see that the Maxwell Hamiltonian commutes with the translation operator, $\left[\mathcal{T}, \bm{\mathcal{H}}\right] = 0$.
In vacuum, the eigen-fields to the Maxwell Hamiltonian will be simple plane waves \mbox{$\bm{f}\sim e^{i\bm{q}\cdot\bm{r}}$}. However, inside a material, the Maxwell Hamiltonian is modulated by a periodic dielectric response, hence the eigen-fields will take a Bloch form \cite{DLJohnson}
\begin{equation}
\bm{f}_{\bm{q}}(\bm{r},\omega) = e^{i\bm{q}\cdot\bm{r}} \bm{u}_{\bm{q}}(\bm{r},\omega),
\end{equation}
where, $\bm{u}_{\bm{q}}$ is the pico-photonic Bloch function, a periodic vector function with the same periodicity as the crystal, and $\bm{q}$ is the photon momentum. $\bm{u}_{\bm{q}}$ can be expanded as a Fourier series of plane waves $e^{i\bm{G}\cdot\bm{r}}$, with $\bm{G}$ being the reciprocal lattice vector
\begin{equation}
\bm{u}_{\bm{q}}(\bm{r},\omega) = \sum_{\bm{G}} \bm{\mathcal{U}}_{\bm{G}}(\bm{q},\omega) e^{i\bm{G}\cdot\bm{r}},  
\end{equation}
where, \mbox{$\bm{\mathcal{U}}_{\bm{G}} = \left[\begin{array}{cc}
\bm{E}_{\bm{G}}     &  \bm{H}_{\bm{G}}
\end{array}\right]^{T}$}.

\subsection{Pico-photonic Eigenvalue Equation}
In a material, the response to an external probe is captured by the displacement field \mbox{$\bm{g}_{\bm{q}}(\bm{r},\omega) = \sum_{\bm{G}} \bm{\mathcal{V}}_{\bm{G}}(\bm{q},\omega) e^{i\bm{G}\cdot\bm{r}}$}, with \mbox{$\bm{\mathcal{V}}_{\bm{G}} = \left[\begin{array}{cc}
\bm{D}_{\bm{G}}     &  \bm{B}_{\bm{G}}
\end{array}\right]^{T}$}. Within a linear response framework, the atomistic displacement field $\bm{\mathcal{V}}_{\bm{G}}$ can be expressed as
\begin{align}
\bm{\mathcal{V}}_{\bm{G}} &= \sum_{\bm{G}'}\bm{\mathcal{R}_{\bm{G}\bm{G}'}}\cdot\bm{\mathcal{U}}_{\bm{G}'},\nonumber\\
\left[\begin{array}{c}
\bm{D}_{\bm{G}}     \\  \bm{B}_{\bm{G}}
\end{array}\right]&= \sum_{\bm{G}'}\left[\begin{array}{cc}
   \varepsilon_T^{\bm{G}\bm{G}'}(\bm{q},\omega)  &\xi_T^{\bm{G}\bm{G}'}(\bm{q},\omega)  \\
   \tau_T^{\bm{G}\bm{G}'}(\bm{q},\omega)  & \mu_T^{\bm{G}\bm{G}'}(\bm{q},\omega)
\end{array}\right]\cdot\left[\begin{array}{c}
\bm{E}_{\bm{G}'}     \\  \bm{H}_{\bm{G}}
\end{array}\right],\nonumber
\end{align}
where, $\bm{\mathcal{R}}_{\bm{G}\bm{G}'}$ is the generalized linear response matrix, which includes permittivity $\varepsilon$, permeability $\mu$, and magneto-electric coupling $\tau$, $\xi$. The component $\bm{\mathcal{R}}_{\bm{G}\bm{G}'}(\bm{q}, \omega)$ can be thought of as the linear response observed in a given material at a field point $\bm{q}+\bm{G}'$ (in reciprocal lattice space) due to a perturbation at the source point $\bm{q}+\bm{G}$. This form can now be substituted into Eq.~(\ref{eq:EMhamiltonian}), and the Maxwell Hamiltonian equation in matter given by
\begin{align}\label{eq:atomisticEMHamiltonian}
\bm{\mathcal{H}}(\bm{q}+\bm{G})\cdot&\left[\begin{array}{c}
     \bm{E}_{\bm{G}}  \\
      \bm{H}_{\bm{G}} 
\end{array}\right] \nonumber\\&= \frac{\omega}{c}\sum_{\bm{G}'}\bm{\mathcal{R}}_{\bm{G}\bm{G}'}(\bm{q}, \omega)\cdot\left[\begin{array}{c}
     \bm{E}_{\bm{G}'}  \\
      \bm{H}_{\bm{G}'} 
\end{array}\right].
\end{align}
$\bm{\mathcal{R}}_{\bm{G}\bm{G}'}$ should build in the space group symmetry of the Brillouin zone, and $\bm{G}, \bm{G}' \neq 0$ terms in the matrix encodes the inhomogeneity due to the microscopic response of the electrons (the local fields). The above Hamiltonian equation depends nonlinearly on the eigenvalue $\omega$ due to the response matrix $\bm{\mathcal{R}}_{\bm{G}\bm{G}'}(\bm{q}, \omega)$. Such class of equations are known as the nonlinear eigenvalue problem. Solutions to this generalized nonlinear eigenvalue problem results in the atomistic electrodynamic dispersion of a material that represents transverse photon interaction in a material system. 

We note that the above Maxwell Hamiltonian equation of motion is based on the plane wave expansion, whose solutions result in the atomistic electrodynamic dispersion. Similarly, it is well known that the electronic bandstructure of a material can be determined by the plane wave expansion of the Schr\"odinger Hamiltonian of the form \cite{CohenBergstresserEPA}
\begin{align}
\sum_{\bm{G}'}\Bigg[\frac{\hbar^2\left|\bm{q}+\bm{G}\right|^2}{2m}\delta_{\bm{G}\bm{G}'} + V(\bm{G}-\bm{G}')&\Bigg]U(\bm{G}') \nonumber\\ &= E\,U(\bm{G}),
\end{align}
and the corresponding electronic wavefunction will be of the form \mbox{$\psi(\bm{r}) = e^{i\bm{k}\cdot\bm{r}}\sum_{\bm{G}}U(\bm{G}) e^{i\bm{G}\cdot\bm{r}}$}.
Hence, the burden of determining the electronic bandstructure of a material falls upon the accurate determination of the pseudopotential coefficients $V(\bm{G}-\bm{G}')$. In a similar manner, one needs to obtain the response matrix $\bm{\mathcal{R}}_{\bm{G}\bm{G}'}(\bm{q}, \omega)$ to deduce the atomistic electrodynamic dispersion.  In Si, only the atomistic dielectric function $\varepsilon_T^{\bm{G}\bm{G}'}$ has considerable contributions, $\mu_T = 1$, and $\xi_T = \tau_T = 0$. In Sec.~\ref{sec:isotropicmodel},  $\varepsilon_T^{\bm{G}\bm{G}'}(\bm{q}, \omega)$ and the atomistic electrodynamic dispersion of Si are obtained within an isotropic nearly-free electron model.   

\subsection{Pico-plasmonic Dispersion}
Here, we go beyond the well known definition of nanoscale plasmons and epsilon-near-zero materials which uses the macroscopic response of matter. We show that the atomistic electrodynamic theory reveals a dispersion relation that embodies the symmetries of the underlying lattice. A plasmon is a self-sustained charge oscillation induced by a longitudinal electric field without the introduction of external charge densities. Since the longitudinal field is purely determined by the scalar potential, from Eq.~(\ref{eq:vextvrelation}), we see that the condition for sustained plasma excitation in a material is given by
\begin{equation}\label{eq:atomisticplasmoncondition}
\det\left[ \varepsilon_L^{\bm{G}\bm{G}'}\left(\bm{q},\omega\right) \right]  = 0. 
\end{equation}
Using the above relation, we can obtain the eigenfrequencies $\omega$ for a fixed $\bm{q}$. Hence, solving this equation one can obtain the atomistic plasmon dispersion of the material. In the continuum limit, we obtain the standard relation
\begin{equation}
\varepsilon_M\left(\bm{q},\omega\right)  = 0,   
\end{equation}
where the macroscopic dielectric function $\varepsilon_M$ is defined as
\begin{equation}\label{eq:macrodielectriclongitudinal}
\varepsilon_M (\omega)= \lim_{\bm{q}\rightarrow 0} \frac{1}{\left(\varepsilon_L^{\bm{G}\bm{G}'}\right)^{-1}_{00}},  
\end{equation}
where, $\left(\varepsilon_L^{\bm{G}\bm{G}'}\right)^{-1}_{00}$ is the first diagonal component of the inverse longitudinal dielectric matrix. Inverse operation indirectly includes the off-diagonal local-field effect contributions. Alternatively, in literature, the plasmon dispersion is determined by identifying the peaks of the energy loss function 
\begin{align}
L(\bm{q},\omega) &= -{\rm Im}\,\varepsilon^{-1}_M\left(\bm{q}, \omega\right), \nonumber\\&= \frac{\varepsilon_2\left(\bm{q}, \omega\right)}{\left[\varepsilon_1\left(\bm{q}, \omega\right)\right]^2+\left[\varepsilon_2\left(\bm{q}, \omega\right)\right]^2},
\end{align} 
where we have taken $\varepsilon_M = \varepsilon_1 + i \varepsilon_2 $.
This is known as the experimental definition of the plasmon dispersion \cite{Farid_plasmondispersion}. At the plasmon frequency $\omega_p$, $\varepsilon_1 (\bm{q}, \omega_p)\approx 0$ and  the damping factor $\varepsilon_2$ is small, so that we observe peaks in the energy loss spectrum \cite{Shekar_MEELS, Shekhar:18}. However, we note that Eq.~(\ref{eq:atomisticplasmoncondition}) provides the most general theoretical relation to obtain the atomistic plasmon dispersion of a material 
\cite{Reiter_plasmondispersion}.

\section{Application to Silicon}\label{sec:isotropicmodel}
Silicon has the diamond cubic crystal structure (Fig.~\ref{fig:bz}(a)) and the first Brillouin zone has the shape of a truncated octahedron (Fig.~\ref{fig:bz}(b)). It has been shown earlier \cite{WalterCohen_wave, waltercohen_wavefreq} that the momentum dependent dielectric function in diamond-type materials is insensitive to the direction of $\bm{q}$. Hence, we can replace the truncated octahedron shape (Fig.~\ref{fig:bz}(b)) of the first Brilloin zone by a sphere (Fig.~\ref{fig:bz}(d)) and obtain the dielectric properties through an isotropic model. Moreover, dielectric screening is not sensitive to the details of the bandstructure since it involves all the valence electrons in the material \cite{Dpenn}. We show that the results obtained through an isotropic nearly-free electron bandstructure agrees well with the exact bandstructure models for Si.   

The nearly-free electron model employed here was first introduced by Penn \cite{Dpenn}. This model allows for the formation of standing waves at the Brillouin zone boundaries and accounts for the Umklapp processes \cite{Auluck_wavevector}. In this scheme, the eigen-energy and wavefunctions of an electron is given by 
\begin{align}
E_{\bm{k}}^{\pm} &= \frac{1}{2}\left[E_{\bm{k}}^0+E_{\bm{k}'}^{0}\pm\sqrt{\left(E_{\bm{k}}^{0}-E_{\bm{k}'}^{0}\right)^2+E_g^2} \right],\nonumber\\
\psi_{\bm{k}}^{\pm} &= \frac{\left(e^{i\bm{k}\cdot\bm{r}}+\alpha_{\bm{k}}^{\pm}e^{i\bm{k}'\cdot\bm{r}}\right)}{\sqrt{1+\left(\alpha_{\bm{k}}^{\pm}\right)^2}},
\end{align}
where, 
\begin{align}
\alpha_{\bm{k}}^{\pm} &= \frac{E_g}{2\left(E_{\bm{k}}^\pm - E_{\bm{k}'}^0\right)}, \nonumber\\
E_{\bm{k}}^0 &= \frac{\hbar^2 k^2}{2m}, \nonumber\\
\bm{k}' &= \bm{k}-\bm{G}_1,\nonumber
\end{align}
$\bm{G}_1 = 2k_f \hat{k}$, $k_f$ is the valence Fermi wavevector, and $E_g$ is the bandgap of the material. Superscripts $+$ and $-$ represents $k > k_f$ (conduction) and $k < k_f$ (valence) bands, respectively. Experimentally measured valence electron density for Si is $n_0 = 0.19\,e^{-}/\angstrom^3$. Now consider a free electron solid with the same density. This will form a Fermi sphere in momentum space. According to Sommerfeld theory \cite{Ckittel_book}, the corresponding valence Fermi wavevector in Si is \mbox{$k_f = (3\pi^2 n_0)^{1/3} = 1.78\,\angstrom^{-1}$}. This will form the fully occupied valence band. An additional conduction band with bandgap $E_g$ is constructed to reflect the semiconducting nature of Si. Wavefunction components with wavevector $\bm{k}' = \bm{k}-\bm{G}_1$ facilitates the Umklapp process. For a given photon momentum $\bm{q}$, $\bm{k} \rightarrow \bm{k}+\bm{q}$ indicates the normal process and $\bm{k} \rightarrow \bm{k}+\bm{q}+\bm{G}_1$ is the Umklapp process.   

In Fig.~\ref{fig:bz}(c) \& (e), we have plotted the exact bandstructure and the  isotropic nearly-free electron bandstructure of Si considered here, respectively. The nearly-free electron bandstructure can be thought of as an isotropic symmetric expansion of the electronic bandstructure around the high-symmetric $\Gamma$ point. This model can reproduce the experimentally observed dielectric properties of silicon (see supplementary information).  
\begin{figure*}
    \centering
    \includegraphics[width = 6in]{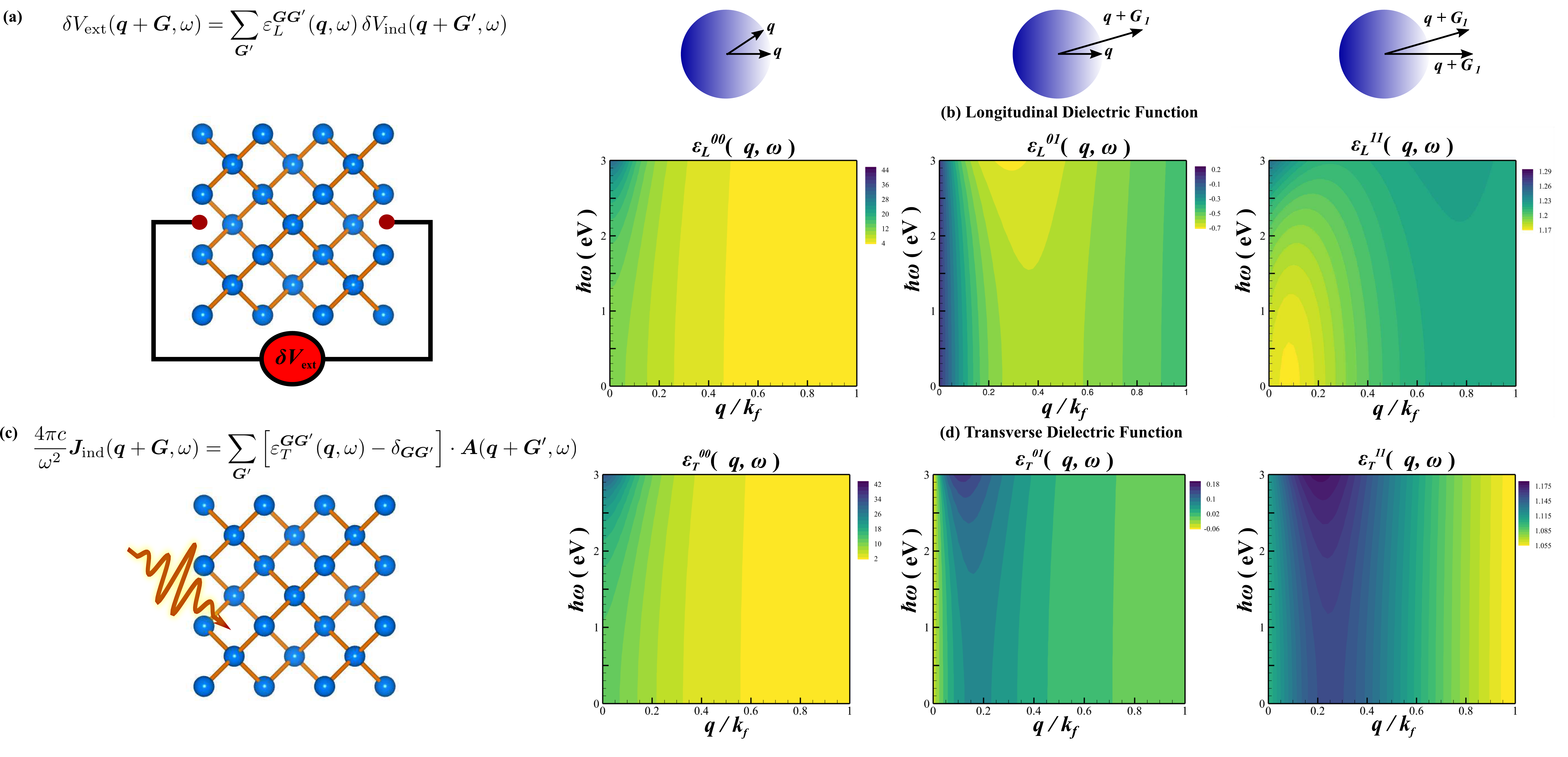}
    \caption{(a) Within the linear response theory, induced potential $\delta V_{\rm ind} (r,t)$ in a material due to an external potential $\delta V_{ext} (r,t)$ can be expressed in terms of the longitudinal dielectric function $\varepsilon_L^{\bm{G}\bm{G}'}$. (c) The transverse dielectric function $\varepsilon_T^{\bm{G}\bm{G}'}$ determine the linear response of a material to a transverse electromagnetic pulse. Contour plots of (b) longitudinal (d) transverse dielectric function components for silicon are displayed as a function of frequency $\omega$ and wavevector $q$.}
    \label{fig:2Ddielectricplot}
\end{figure*}

We will now proceed to obtain the longitudinal and transverse dielectric function of Si using this model. Through inspection, we see that for either case, within this model only the dielectric matrix elements corresponding to $\bm{G} = 0$ and $\bm{G}_1 = 2k_f \hat{k}$ are non-zero. All higher order elements corresponding to the reciprocal lattice vectors vanish.
\begin{figure*}
    \centering
    \includegraphics[width = 4in]{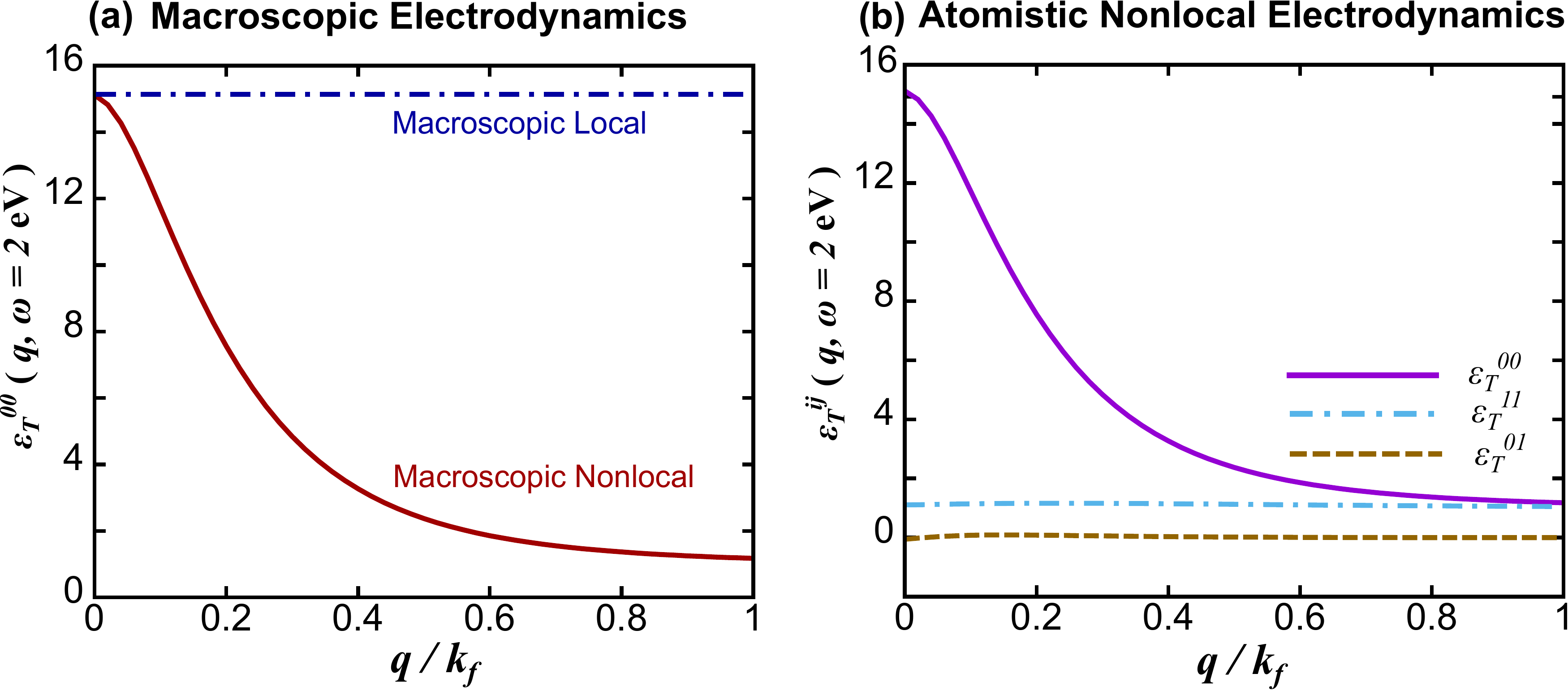}
    \caption{The transverse dielectric function $\varepsilon_T^{ij}(q, \hbar \omega = 2\,$eV$)$ is plotted as a function of wavevector $q$. (a) In a macroscopic local theory, the dielectric component $\varepsilon_T^{00}$ is independent of $q$. Whereas, in case of the macroscopic nonlocal framework, only the $\varepsilon_T^{00}(q, \omega)$ component is considered, and the local-field effects are neglected. (b) In an atomistic nonlocal theory, $\varepsilon_T^{00}, \varepsilon_T^{01}$, and $\varepsilon_T^{11}$ components have significant variation with $q$ and contribute to the overall dielectric response of the material.}
    \label{fig:dielectricomparisontransverse}
\end{figure*}

Typically, the dielectric function of a material is considered to be only a function of $\omega$. In Fig.~\ref{fig:2Ddielectricplot}, we observe a family of curves dependent on the wavevector $q$ even at a fixed $\omega$. Moreover, the evolution of longitudinal and transverse dielectric function are found to be inequivalent at $\bm{q} \neq 0$. We note that \mbox{$\varepsilon_{L}^{ij}(q \neq 0, \omega)$} and \mbox{$\varepsilon_{T}^{ij}(q \neq 0, \omega)$} represent the atomistic nonlocal contributions to the dielectric properties. $\varepsilon_{L}^{01}$ and $\varepsilon_{T}^{01}$ (corresponding to \mbox{$\bm{G} = 0, \bm{G}' = \bm{G}_1$}), $\varepsilon_{L}^{11}$ and $\varepsilon_{T}^{11}$ (corresponding to \mbox{$\bm{G} = \bm{G}' = \bm{G}_1$}) are due to the local-field effects. $\varepsilon_{L}^{00}(q \neq 0, \omega)$ and $\varepsilon_{T}^{00}(q \neq 0, \omega)$ determine the dielectric response of a material at a source and field point on the sphere of radius $q$. Since we have considered an isotropic electron model, this dielectric response is identical at all points on this sphere. Whereas, $\varepsilon_{L,T}^{01}(q \neq 0, \omega)$ determines the material response at a field point $\bm{q} + \bm{G}_1$ from a source point $\bm{q}$ in momentum space, and $\varepsilon_{L,T}^{11}(q \neq 0, \omega)$ is the dielectric response from a source and field point both at $\bm{q}+\bm{G}_1$. These scenarios are pictorially depicted in Fig.~\ref{fig:2Ddielectricplot}.

In literature, typically only the longitudinal dielectric function in the long-wavelength limit $\varepsilon_L^{00}(\bm{q} = 0, \omega)$ is calculated and used to obtain all dielectric properties of the material. Our calculations show that at finite momentum $(\bm{q} \neq 0)$, transverse and longitudinal dielectric function are inequivalent, and the higher-order components have significant contributions to the dielectric properties even at $\omega = 0$. In our analysis we neglect the damping factor contributions in the dielectric response. In the next section, we show that the local-field contributions lead to an additional anomalous band formation in the atomistic electrodynamic dispersion of Si. 

\section{Anomalous Band in the Forbidden Gap}\label{sec:atomisticEMbandstructure}
\begin{figure*}[t]
    \centering
    \includegraphics[width = 6.5in]{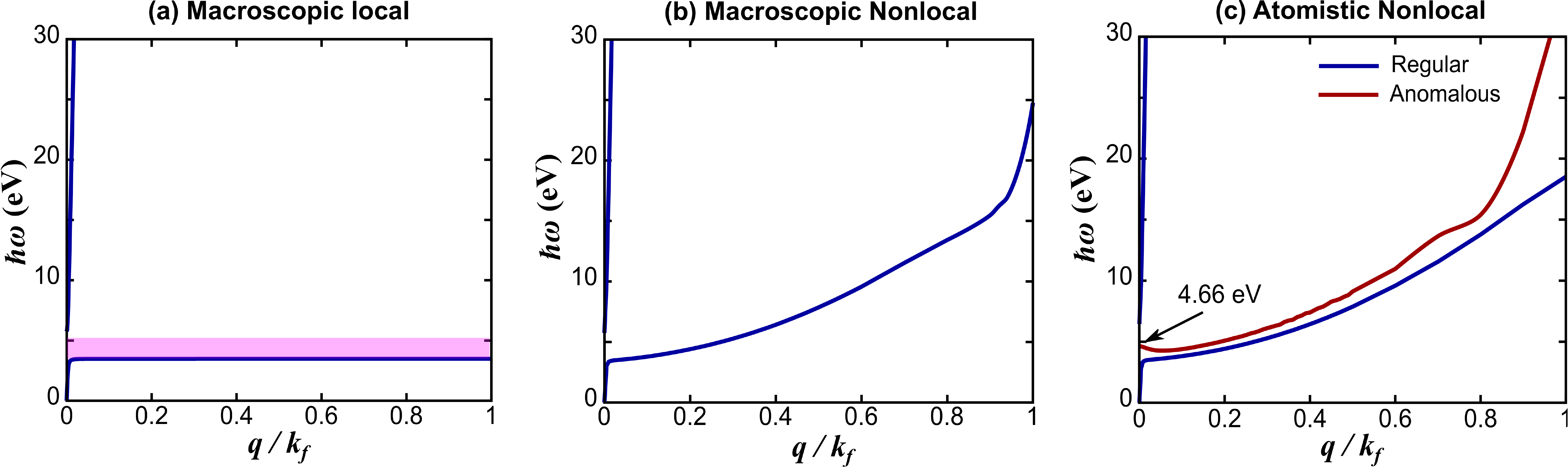}
    \caption{Atomistic electrodynamic dispersion of silicon is plotted as a function of momentum obtained through (a) a macroscopic local electromagnetic theory, (b) a macroscopic nonlocal theory, (c) an atomistic nonlocal electrodynamic theory. In the later case, we observe the emergence of an anomalous band in the electrodynamic dispersion.}
    \label{fig:embandstructure}
\end{figure*}

In this section, we apply the Maxwell Hamiltonian framework described in Sec.~\ref{sec:atomisticEMtheory} to obtain the atomistic electrodynamic $\omega-q$ dispersion in Si. We show the existence of anomalous atomistic waves in the bandgap of silicon. We also directly compare with existing theories to recover well-known waves and also prove that these new waves are the result of atomistic electrodynamics. 

Components of the transverse dielectric function are obtained through the isotropic nearly free-electron model (see Fig.~\ref{fig:2Ddielectricplot}(b)). Below, we outline two theoretical approaches that we use as a comparison to our atomistic nonlocal electrodynamic theory.  

Macroscopic local theory: In a macroscopic local electrodynamic theory, the dielectric function is only dependent on the frequency while the local-field effects are ignored. Hence, only $\varepsilon_T^{00}(q = 0, \omega)$ contributes to the dielectric properties of the material. For a given frequency, $\varepsilon_T^{00}$ is considered constant across the momentum range (Fig.~\ref{fig:dielectricomparisontransverse}(a)). In a macroscopic theory, transverse electromagnetic waves satisfy the continuum relation
\begin{equation}
q^2 = \varepsilon_T^{00}(q = 0, \omega)\frac{\omega^2}{c^2}.
\end{equation}
Solution to the above equation results in the electrodynamic dispersion shown in Fig.~\ref{fig:embandstructure}(a). We observe the light cone behavior retained for small $q$ values. A bandgap is observed in the spectrum corresponds to the region $\varepsilon_T^{00} < 0$. At large $q$ values photons are localized (zero slope of the band) consistent with the local dielectric response considered here. 

Macroscopic nonlocal theory: Dielectric function behavior at $q \neq 0$ determines the nonlocal response of the material. Hence, in case of a macrosocpic nonlocal theory, we consider the dielectric response to be $\varepsilon_T^{00}(q, \omega)$ and the local-field effects are again neglected. In Fig.~\ref{fig:dielectricomparisontransverse}(a), we compare the dielectric function behavior considered within a macroscopic local and macroscopic nonlocal theory. With increase in momentum, we observe a decaying behavior in the dielectric function $\varepsilon_T^{00}(q, \omega)$ at any given frequency. Within this framework, transverse electromagnetic waves satisfy the continuum relation
\begin{equation}
q^2 = \varepsilon_T^{00}(q, \omega)\frac{\omega^2}{c^2}.
\end{equation}  
In Fig.~\ref{fig:embandstructure}(b), we observe that at large $q$ values, electrodynamic bands have a finite slope due to nonlocal response of the material. We call the dispersion curves observed in Fig.~\ref{fig:embandstructure}(a) and (b) through a macroscopic theory as the regular bands. 

Atomistic nonlocal electrodynamic theory: In Fig.~\ref{fig:dielectricomparisontransverse}, we compare the dielectric response in a macroscopic and an atomistic electrodynamic theory. In a macroscopic theory, local-field effects are neglected. Hence, the dielectric response has a single component. However, we see that the higher-order dielectric components $\varepsilon_T^{01}, \varepsilon_T^{11}$ have small but non-negligible contributions to the overall dielectric response of the material. The generalized nonlinear pico-photonic eigenvalue problem in Eq.~(\ref{eq:atomisticEMHamiltonian}) is solved to obtain the atomistic electrodynamic dispersion (see supplementary information). 

In Fig.~\ref{fig:embandstructure}(c), we see that along with regular bands, an anomalous band is also observed in the dispersion. This anomalous band is absent if we treat the problem using macroscopic local or macroscopic nonlocal electrodynamic frameworks. Hence, the anomalous band is a direct consequence of the inclusion of local-field effects in Si. Even at $q = 0$, the anomalous band has a finite frequency. This is in stark contrast with the regular band, whose frequency vanishes at $q =0$. In classical optical theories one would consider this regime to be perfectly metallic where the propagation of light is forbidden. However, from Fig.~\ref{fig:embandstructure}(c), we see that the light can propagate through silicon in the pico-photonics regime. 
\begin{figure}[h]
    \centering
    \includegraphics[width = 2.5in]{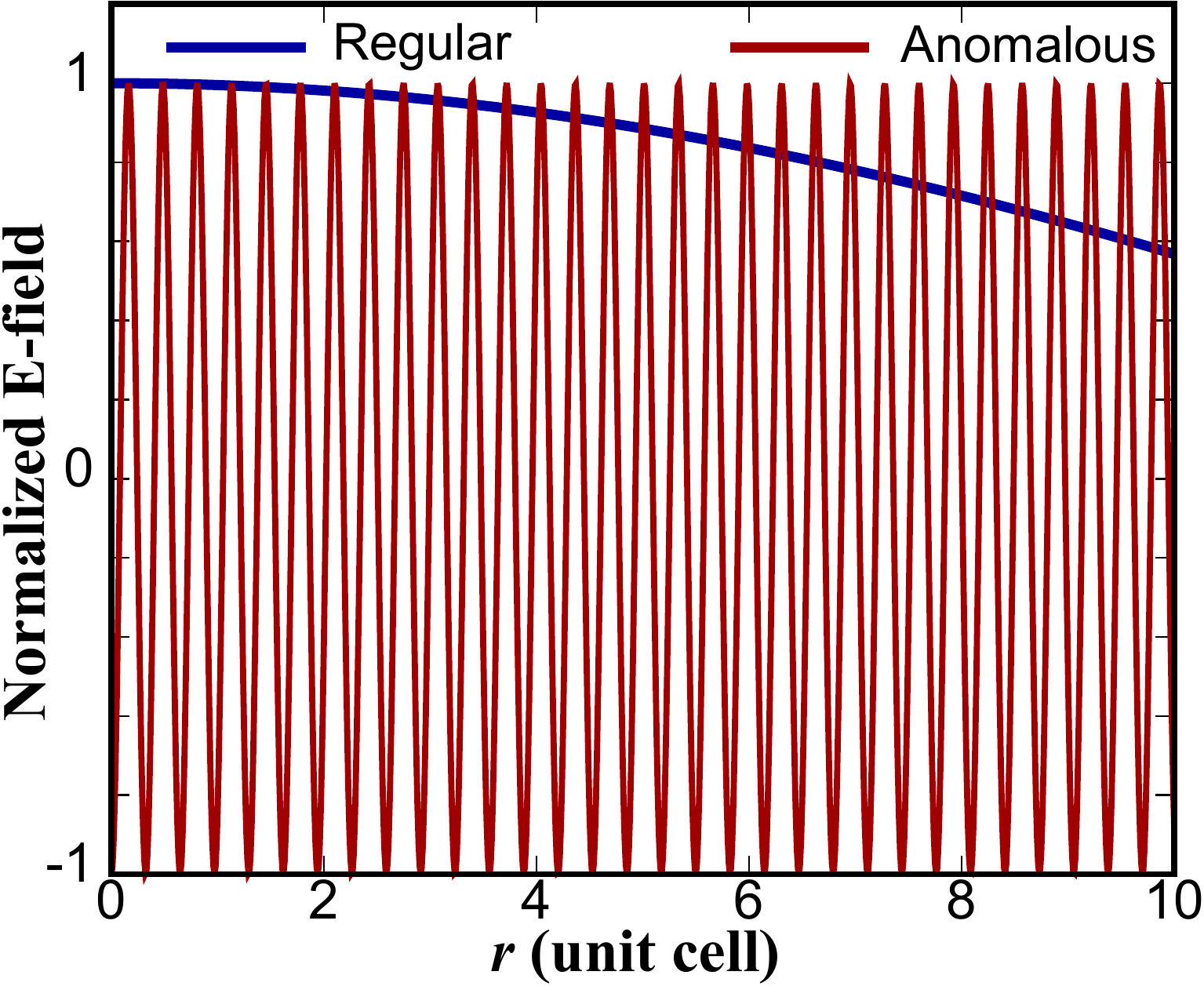}
    \caption{Normalized electric field is plotted at $q = 0.178\,$nm$^{-1}$  for the regular and anomalous bands. Here, the regular band (blue curve) has the wavelength $\lambda = 35.30\,$nm. Where as, the anomalous band (red curve) has $\lambda = 0.18\,$nm, in the pico-photonics regime.}
    \label{fig:efieldvariation}
\end{figure}
 
 In Fig.~\ref{fig:efieldvariation}, we plot the normalized electromagnetic field at $q = 0.178\,$ nm$^{-1}$ in Si hosted by the regular and anomalous band. Across the momentum, the regular band has wavelengths in nano-meters, whereas the anomalous band has sub-nm wavelengths. The lattice constant of a silicon unit cell is $0.543\,$nm, hence electromagnetic waves in the anomalous band are found to be highly oscillatory within a unit cell, leading into the pico-photonics regime. 

For completeness, in Appendix~\ref{appdx:plasmonbandstructure}, we have calculated the atomistic plasmon dispersion of Si obtained within the isotropic nearly-free electron model. In Fig.~\ref{fig:plasmondispersion}, we observe that the atomistic nonlocal and macroscopic nonlocal theory results in a nearly identical plasmon dispersion across the momentum range. 
\begin{figure*}
    \centering
    \includegraphics[width = 5in]{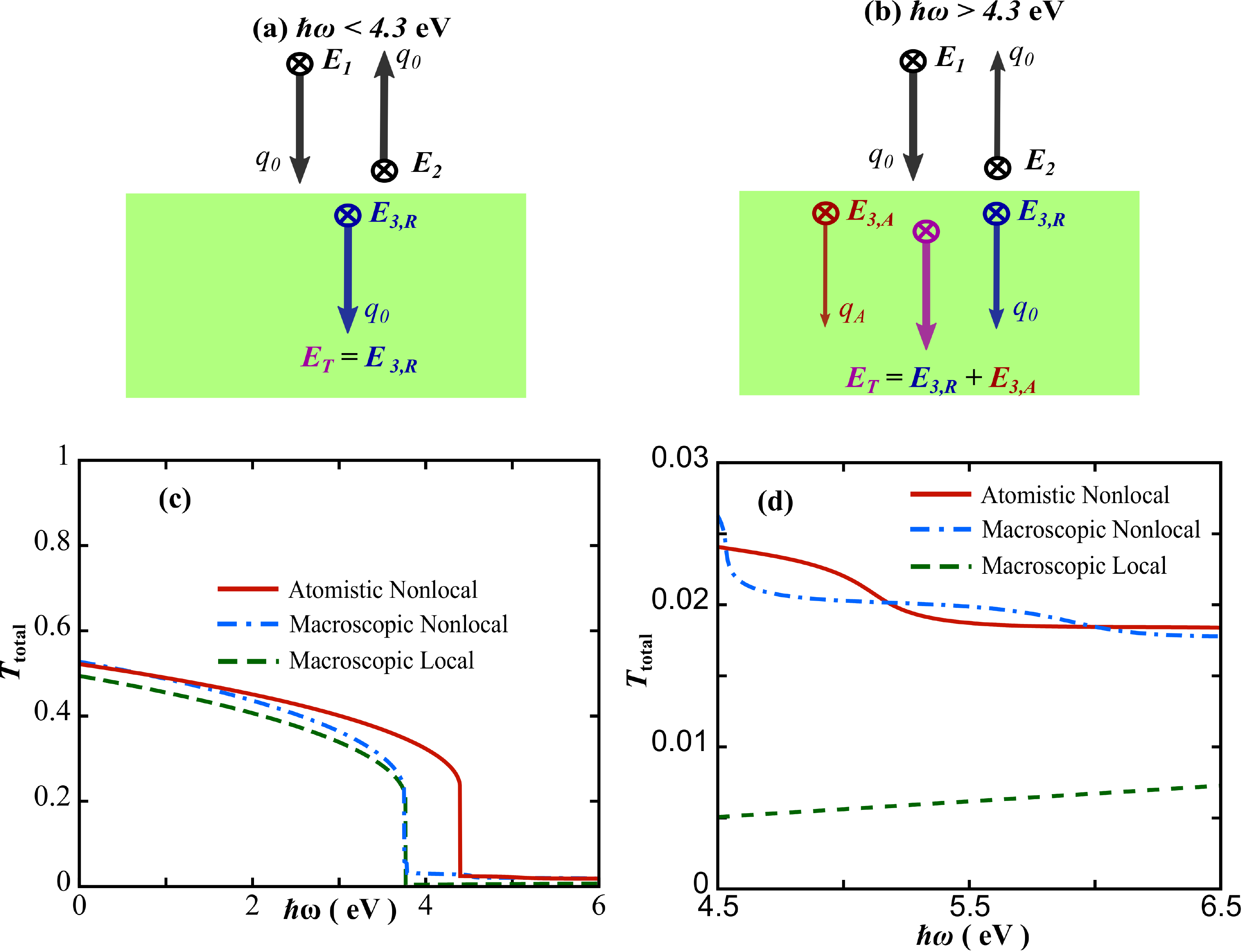}
    \caption{(a) For energies $\hbar \omega < 4.3\,$eV, only the regular band contributes to the total transmission in a silicon block. (b) For energies $\hbar \omega > 4.3\,$eV, both regular and anomalous bands are excited by an electromagnetic wave incident on a silicon block. (c) The total transmission coefficient $T_{\rm total}$ at normal incidence is plotted as a function of energy $\hbar \omega$. (d) The total transmission coefficient at normal incidence is plotted as a function of energy in the deep ultraviolet regime, where both regular and anomalous bands contribute to the total transmission. We observe a clear difference in the behavior of $T_{\rm total}$ obtained using the atomistic nonlocal theory while compared to the macroscopic theory. This difference is attributed to the contributions from the anomalous band.}
    \label{fig:transmission}
\end{figure*}
\subsection{Experimental probe of anomalous atomistic waves}
We propose an experiment to probe the atomistic pico-photonic dispersion relation in silicon. Consider an electromagnetic wave incident normally on an Si block (see Fig.~\ref{fig:transmission}(a)). Experimentally, one can control the energy, whereas the momentum within the crystal is determined by the atomistic electrodynamic dispersion. We calculate the transmission coefficient at two different energy ranges, (a) \mbox{$\hbar\omega < 4.3\,$eV}, and (b) \mbox{$\hbar\omega > 4.3\,$eV}. For energies $\hbar\omega < 4.3\,$eV, only the regular band is excited. Hence the total transmission in Si block will have contributions only from the regular band (Fig.~\ref{fig:transmission}(a)). From Fig.~\ref{fig:embandstructure}(c) we see that for energies $\hbar\omega > 4.3\,$eV both regular and anomalous bands are excited with two distinct momentum. Hence, the total transmission should include additional terms from the interference effects due to field contributions of the anomalous band (see supplementary information for calculation details). 

In Fig.~\ref{fig:transmission}(c), we have plotted the total transmission coefficient calculated using the macroscopic local, macroscopic nonlocal and the atomistic nonlocal theory. We observe that all three calculations have similar results for low energies. However, in deep ultraviolet regime (DUV) $(\hbar\omega > 4.3$\,eV$)$, as shown in Fig.~\ref{fig:transmission}(d), the atomistic nonlocal theory displays a significantly different behavior from that of macroscopic theories due to interference between regular and anomalous bands. We note that in the DUV regime, the anomalous band generates additional electromagnetic energies at a given frequency of light. This additional energy contribution is reflected in the behavior of total transmission coefficient determined through our atomistic nonlocal theory.

Experimentally, one can measure the total transmission in the range \mbox{$4.5\,$eV$ < \hbar \omega < 6.5\,$eV} by shining an ultraviolet light on a silicon block. The total transmission can be accurately measured using high sensitivity single photon detectors. Difference in the measured total transmission coefficient to that of macroscopic electrodynamic calculations should reveal the existence of anomalous bands in the atomistic electrodynamic dispersion. In our calculations we have not considered the phonon mediated inter-band transitions in silicon. However, such transitions can be suppressed in low-temperature experiments. 
\section{Conclusions}\label{sec:conclusions}
We have developed the atomistic nonlocal electrodynamic theory of matter through a Maxwell Hamiltonian framework. We introduced the atomistic transverse dielectric tensor which determines the linear response of a material to a transverse electromagnetic probe. The electrodynamics of matter is considered in a new light through the Maxwell Hamiltonian, which captures the \mbox{spin-1} nature of photons. Through this formulation, we have discovered anomalous waves in the atomistic electrodynamic dispersion of silicon. Local-field effects included in the Maxwell Hamiltonian are essential to obtain the anomalous waves in the atomistic electrodynamic dispersion. These waves are highly oscillatory within a unit cell and have sub-nm wavelengths in the pico-photonics regime. The anomalous wave generates an additional electromagnetic energy contribution which was previously unaccounted for. Experimental signatures for this additional electromagnetic energy contribution can be deduced from the total transmission coefficient in the deep ultraviolet regime. We note that the frequencies corresponding to the anomalous waves are forbidden in a macroscopic local model, and are a signature of our quantum theory of atomic polarization developed here. 

Our findings demonstrate that natural media can itself host several interesting electrodynamic phases. As such, the electrodynamic phases we discussed here are properties of matter itself and are not related to some form of macroscopic engineering. In this study we considered Si as a prototype material. Ge, AlSb, ZnSe, GaAs, GaP, InP, ZnS, ZnTe, CdTe are all expected to display the anomalous band, since all these material systems have the same crystal symmetry as Si. We expect to observe anomalous waves in many other natural materials.  

Results presented here brings forth the importance of the atomistic electrodynamic phases of matter, and the immediate need to develop first-principles based atomistic nonlocal electrodynamics of matter to obtain the atomistic electrodynamic dispersion of natural materials. We envision the development of pico-photonic electrodynamic density functional theory (PED-DFT) for photons hosted by matter to reveal new effects connected to the atomistic electrodynamic dispersion. Our analysis provides the fist step towards the discovery of topological photonic properties in natural materials.
\section{Acknowledgements}
This work was supported by the Defense Advanced Research Projects Agency (DARPA) under Quest for Undiscovered Energy Storage and Thrust (QUEST) program. 
\appendix
\section{Maxwell Hamiltonian in Free Space}\label{appdx:MaxwellHamiltonian}
In this appendix, we derive the Maxwell Hamiltonian discussed in Sec.~\ref{sec:atomisticEMtheory}. The Maxwell's equations (in Gaussian units) are given by
\begin{align}
\nabla \cdot \bm{E} = 4\pi\rho, \quad& \nabla \cdot \bm{B} = 0,\nonumber\\
\nabla \times \bm{E} = -\frac{1}{c}\frac{\partial \bm{B}}{\partial t}, \quad& \nabla \times \bm{B} = \frac{1}{c}\frac{\partial \bm{E}}{\partial t}+\frac{4\pi}{c}\bm{J}.
\end{align}
Along with the above equations, the charge density and current density have to satisfy the continuity equation
\begin{equation}
\nabla\cdot\bm{J}+\frac{\partial \rho}{\partial t} = 0. 
\end{equation}
Fields can be expressed in terms of the scalar and vector potentials of the form
\begin{align}
\bm{E} = -\nabla V -\frac{1}{c}\frac{\partial\bm{A}}{\partial t}, \quad \bm{B} = \nabla\times \bm{A}. 
\end{align}
These potentials satisfy the gauge transformations
\begin{align}
V \rightarrow&\ V -\frac{1}{c}\frac{\partial \Xi}{\partial t}, \nonumber\\
\bm{A} \rightarrow&\ \bm{A} +\nabla\Xi,      
\end{align}
where $\Xi$ is the gauge function. 
It is convenient to decompose the electric field in terms of longitudinal and transverse components, given by
\begin{equation}
\bm{E}(\bm{r},t) = \bm{E}_L(\bm{r},t) + \bm{E}_T(\bm{r},t),   
\end{equation}
where, $\nabla \cdot \bm{E}_T(\bm{r},t) = 0$ and $\nabla \times \bm{E}_L(\bm{r},t) = 0$. Notice that the magnetic field will have only transverse component due to zero-divergence condition. With this decomposition, one can write
\begin{align}\label{eq:maxwelllt}
\nabla \cdot \bm{E}_L = 4\pi\rho, \quad& \nabla \cdot \bm{B}_T = 0,\nonumber\\
\nabla \times \bm{E}_T = -\frac{1}{c}\,\frac{\partial \bm{B}_T}{\partial t} , \quad& \nabla \times \bm{B}_T = \frac{1}{c}\frac{\partial \bm{E}}{\partial t}+\frac{4\pi}{c}\bm{J},
\end{align}
and the corresponding gauge transformations are given by
\begin{align}
V \rightarrow&\ V -\frac{1}{c}\frac{\partial \Xi}{\partial t}, \nonumber\\
\bm{A}_L \rightarrow&\ \bm{A}_L +\nabla\Xi,\\
\bm{A}_T \rightarrow&\ \bm{A}_T .\nonumber
\end{align}
We can choose the gauge function $\Xi$ such that $\bm{A}_L = 0$. Hence
\begin{align}
\bm{E}_L = -\nabla V,\quad \bm{E}_T = -\frac{1}{c}\frac{\partial\bm{A}_T}{\partial t}, \quad \bm{B} = \nabla\times \bm{A}_T.
\end{align}
Hence, the longitudinal electric field $\bm{E}_L$ is purely determined by the scalar potential. The Maxwell Hamiltonian is related to the transverse part of the electromagnetic fields. We first consider the Amp\'ere–Maxwell equation given by
\begin{align}
\nabla \times \bm{B}_T &= \frac{1}{c}\frac{\partial \bm{E}}{\partial t}+\frac{4\pi}{c}\bm{J},\nonumber\\
&= \frac{1}{c}\frac{\partial \bm{E}_L}{\partial t}+\frac{1}{c}\frac{\partial \bm{E}_T}{\partial t}+\frac{4\pi}{c}\bm{J}_L+\frac{4\pi}{c}\bm{J}_T.\label{eq:transamperelaw}
\end{align}
We can show that 
\begin{align}
\nabla\cdot\frac{\partial \bm{E}_L}{\partial t} &= 4\pi \frac{\partial \rho}{\partial t},\nonumber\\
&= -4\pi \nabla\cdot \bm{J}_L.\nonumber    
\end{align}
Hence,
\begin{equation}
\frac{\partial \bm{E}_L}{\partial t} = -4\pi\bm{J}_L.
\end{equation}
Using this relation, we can simplify Eq.~(\ref{eq:transamperelaw}) as
\begin{equation}\label{eq:transamperesim}
\nabla \times \bm{B}_T = \frac{1}{c}\frac{\partial \bm{E}_T}{\partial t}+\frac{4\pi}{c}\bm{J}_T.  
\end{equation}
We are interested in the response of a bulk material. Therefore, it is convenient to represent the induced charges and current in terms of
the polarization $\bm{P}$ and magnetization density $\bm{M}$,
\begin{align}
\rho = -\nabla\cdot\bm{P}_L,\ \bm{J} = \frac{\partial \bm{P}}{\partial t}+c\,\nabla\times\bm{M}. 
\end{align}
The equations of motion in terms of the displace fields $\bm{D} = \bm{E} + 4\pi\bm{P}$, and $\bm{H} = \bm{B}-4\pi \bm{M}$ are given by
\begin{align}
\nabla \times \bm{E}_T = -\frac{1}{c}\frac{\partial \bm{B}_T}{\partial t},  \ \nabla \times \bm{H}_T = \frac{1}{c}\frac{\partial \bm{D}_T}{\partial t}.     
\end{align}
Hamiltonian form presented in Eq.~(\ref{eq:EMhamiltonian}) immediately follows if we define \mbox{$\bm{f} = \left[\begin{array}{cc}
     \bm{E}_T &
      \bm{H}_T
\end{array}\right]^T$}, and \mbox{$\bm{g} = \left[\begin{array}{cc}
     \bm{D}_T &
      \bm{B}_T
\end{array}\right]^T$}.

\section{Atomistic Plasmon dispersion in Silicon}\label{appdx:plasmonbandstructure}
\begin{figure}
    \centering
    \includegraphics[width = 2.5in]{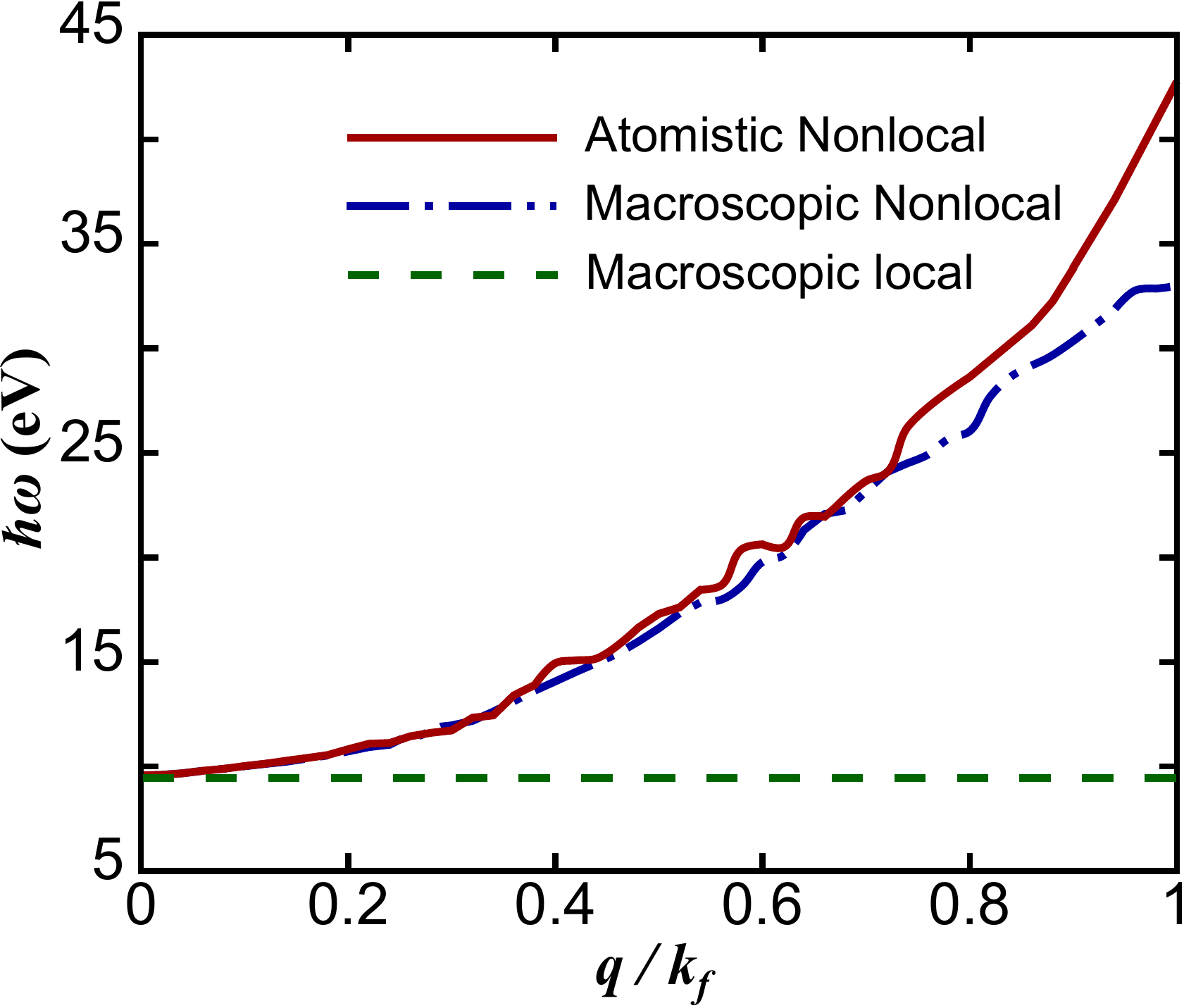}
    \caption{Atomistic plasmon dispersion of silicon is plotted as a function of wavevector $q$. We have compared the results obtained through a macroscopic local, macroscopic nonlocal, and an atomistic nonlocal electrodynamic theory. Macrosocpic nonlocal and atomistic nonlocal electrodynamic theory results in nearly identical plasmon dispersion.}
    \label{fig:plasmondispersion}
\end{figure}

The atomistic plasmon dispersion of Si has been studied both theoretically \cite{Farid_plasmondispersion} and experimentally \cite{Raether_plasmondispersion} previously in literature. For completeness, in this appendix, we present the atomistic plasmon dispersion obtained using the isotropic nearly-free electron model. In Fig.~\ref{fig:2Ddielectricplot}(b), we have displayed $\varepsilon_L^{00}(q, \omega)$, $\varepsilon_L^{01}(q, \omega)$, and $\varepsilon_L^{11}(q, \omega)$ for Si obtained using this model. We can substitute these functions into Eq.~(\ref{eq:atomisticplasmoncondition}) to obtain the plasmon dispersion using the atomistic nonlocal electrodynamic theory (Fig.~\ref{fig:plasmondispersion}). As earlier, we compare the dispersion obtained through the macroscopic local and macroscopic nonlocal theory.

In case of macroscopic local theory, dielectric function $\varepsilon_L^{00} (q = 0, \omega)$ is considered independent $\bm{q}$. Hence, the plasmon dispersion curve is observed to be a straight line with zero slope and intercept given by the zero of $\varepsilon_L^{00} (q = 0, \omega)$. In the macroscopic nonlocal theory, plasmon frequencies are determined by the condition $\varepsilon_L^{00} (q , \omega) = 0$. Both macroscopic nonlocal and atomistic nonlocal theory results in nearly identical plasmon dispersion, diverging slightly only at very large $q$. We note that at $q = 0$, plasmon frequency ($\sim 9.6\,$eV) obtained through isotropic nearly-free electron model slightly underestimates the corresponding experimentally observe value ($\sim 16\,$eV) \cite{Raether_plasmondispersion}.

\bibliography{references}


\pagebreak

\onecolumngrid
\begin{center}
\textbf{Supplementary Information:\\Pico-photonics: Anomalous Atomistic Waves in Silicon}
\end{center}
In this supplementary information, we derive the atomistic dielectric function of a material as a function of both frequency and momentum, including the local-field effects. We show that both the longitudinal and transverse dielectric function can be expressed in terms of energy eigenvalues and the corresponding electronic Bloch functions of a material. Further, we employ a nearly-free electron bandstructure to obtain the atomistic dielectric function of silicon as a function of both frequency and momentum. These dielectric functions are further employed to solve the Maxwell Hamiltonian in silicon. Finally, through the atomistic electrodynamic dispersion, we derive the expression for total transmission coefficient at normal incidence including both regular and anomalous band contributions.

	\section{Longitudinal Dielectric Function}\label{sec:longitudinaldielectric}
	From the linear response theory, induced scalar potential $\delta V_{\rm ind}(\bm{r},t)$ in a material due to an external potential $\delta V_{\rm ext}\left(\bm{r},t\right)$ is given by 
	\begin{align}
		\delta V_{\rm ind}\left(\bm{r},t\right) = \int d\bm{r}' dt'\, \varepsilon_L^{-1}\left(\bm{r}, \bm{r}', t - t'\right) \delta V_{\rm ext}\left(\bm{r}',t'\right),   
	\end{align}
	where $\varepsilon_L$ is the longitudinal dielectric function (density-density correlation function). Equivalently one can write 
	\begin{align}
		\delta V_{\rm ext}\left(\bm{r},t\right) = \int d\bm{r}' dt'\, \varepsilon_L\left(\bm{r}, \bm{r}', t - t'\right) \delta V_{\rm ind}\left(\bm{r}',t'\right). 
	\end{align}
	Given a photon momentum $\bm{q}$ and frequency $\omega$, in the Fourier space we obtain 
	\begin{align}
		\delta V_{\rm ext}\left(\bm{q},\omega\right) = \sum_{\bm{q}'} \varepsilon_L\left(\bm{q}, \bm{q}', \omega\right) \delta V_{\rm ind}\left(\bm{q}',\omega\right).  
	\end{align}
	A pure material system satisfy the translation symmetry, therefore the dielectric function follows the relation
	\begin{align}
		\varepsilon_L\left(\bm{r},\bm{r}', t\right) = 
		\varepsilon_L\left(\bm{r}+\bm{R},\bm{r}'+\bm{R}, t\right),
	\end{align}
	where, $\bm{R}$ is the translation vector in real space. 
	Hence, in the reciprocal space 
	\begin{align}
		\delta V_{\rm ext}\left(\bm{q}+\bm{G},\omega\right) &= \sum_{\bm{G}'} \varepsilon_L\left(\bm{q}+\bm{G}, \bm{q}+\bm{G}', \omega\right) \delta V_{\rm ind}\left(\bm{q}+\bm{G}',\omega\right),\nonumber\\
		&= \sum_{\bm{G}'} \varepsilon_L^{\bm{G}\bm{G}'}\left(\bm{q}, \omega\right) \delta V_{\rm ind}\left(\bm{q}+\bm{G}',\omega\right),\label{eq:vextvrelation}
	\end{align}
	where $\bm{q}$ is constrained within the first Brillouin zone, and $\bm{G},\bm{G}'$ are the reciprocal lattice vectors.
	Again from the linear response theory, the induced charge density $\delta \rho$ is expressed as 
	\begin{align}
		\delta \rho\left(\bm{q}+\bm{G},\omega\right) &= \sum_{\bm{G}'} \chi^{\bm{G}\bm{G}'}\left(\bm{q}, \omega\right) \delta V_{\rm ext}\left(\bm{q}+\bm{G}',\omega\right), \label{eq:chi}\\
		\delta \rho\left(\bm{q}+\bm{G},\omega\right) &= \sum_{\bm{G}'} \chi_0^{\bm{G}\bm{G}'}\left(\bm{q}, \omega\right) \delta V_{\rm ind}\left(\bm{q}+\bm{G}',\omega\right),\label{eq:chi0}
	\end{align}
	where $\chi^{\bm{G}\bm{G}'}$ and $\chi_0^{\bm{G}\bm{G}'}$ are the susceptibility tensors. We can expand $\delta V$ as
	\begin{align}
		\delta V_{\rm ind} &= \delta V_{\rm ext} + \delta V_{H} + \delta V_{\rm xc}, \nonumber\\
		&= \delta V_{\rm ext} +  V_{c}\delta \rho +  K_{\rm xc} \delta \rho, \label{eq:deltav}
	\end{align}
	where $V_c(q) = 4\pi e^2/q^2$ is the bare coulomb potential, and $K_{\rm xc}$ is the term coming from exchange correlation. Using Eqs.~(\ref{eq:chi}), (\ref{eq:chi0}), and (\ref{eq:deltav}), we can obtain the relation
	\begin{equation}
		\frac{1}{\chi} = \frac{1}{\chi_0}-V_c-K_{xc}, 
	\end{equation}
	and the longitudinal dielectric function is given by
	\begin{equation}
		\varepsilon_L = \frac{\chi_0}{\chi} = 1-V_c \chi_0 -K_{xc} \chi_0. \label{eq:longitudinaldielectric}     
	\end{equation}
	We will obtain the expression for $\chi_0$ and hence $\varepsilon_L$ using the Kohn-Sham orbitals. The slater determinant form of the ground state in terms of the Kohn-Sham orbitals is given by
	\begin{equation}
		\psi_0\left(\left\{\bm{r}_l\right\}\right) = \frac{1}{\sqrt{N!}}\left|\overbar{\phi_1}\ \overbar{\phi_2}\ ...\overbar{\phi_N}\right|,
	\end{equation}
	where
	\begin{equation}
		\overbar{\phi_l} = \left(\begin{array}{c}
			\phi_l\left(\bm{r}_1\right)  \\
			\phi_l\left(\bm{r}_2\right) \\
			.\\
			.\\
			.\\
			\phi_l\left(\bm{r}_N\right)
		\end{array}\right). \nonumber    
	\end{equation}
	We note that the Kohn-Sham orbitals satisfy the differential equation \cite{cohen_louie_2016} of the form
	\begin{equation}
		\left\{\frac{p^2}{2m}+{v}(\bm{r})+{v}_{H}(\bm{r})+{v}_{xc}(\bm{r})\right\}\phi_i(\bm{r}) = E_i\,\phi_i(\bm{r}),
	\end{equation}
	where the Hatree-potential term ${v}_H = e^2\int\,d\bm{r}\, \rho(\bm{r}')/\left|\bm{r}-\bm{r}'\right|$ and ${v}_{xc}$ is the exchange correlation potential. We know that these Kohn-Sham orbitals follow the Bloch form
	\begin{align}
		\phi_i(\bm{r}) &= u_n(\bm{r})e^{i\bm{k}\cdot\rm{r}}, \\ \nonumber
		&= \left<\bm{r}| n,\bm{k}\right>e^{i\bm{k}\cdot\rm{r}}.  
	\end{align}
	Here, the index $i$ embed both the band index $n$ and the electron momentum $\bm{k}$ ($i \equiv {\bm{k}, n}$).
	The expectation value of density operator in ground state is given by
	\begin{align}
		\rho\left(\bm{r}\right) &= \left<\psi_0\right|\sum_i \delta\left(\bm{r}-\bm{r}_i\right)\left|\psi_0\right>, \nonumber\\
		&= \sum_{l = 1}^{N} \phi^*_l\left(\bm{r}\right)\phi_l\left(\bm{r}\right).
	\end{align}
	The excited state wavefunction for an electron to transmit from occupied orbital $\phi_i$ to unoccupied orbital $\phi_j$ is denoted as
	\begin{equation}
		\psi_{ij} \left(\left\{\bm{r}_l\right\}\right) = \frac{1}{\sqrt{N!}}\left|\overbar{\phi_1}\ ... \overbar{\phi_{i-1}}\ \overbar{\phi_{j}}\ \overbar{\phi_{i+1}}...\overbar{\phi_N}\right|.   
	\end{equation}
	The material system is subject to an adiabatic perturbation by a potential $\delta V$, and the corresponding Hamiltonian is given by
	\begin{equation}
		H_I = \lim_{\alpha \rightarrow 0} e^{-i\omega t}e^{\alpha t} \sum_i \delta V_{\rm ind}\left(\bm{r}_i\right) = \int d\bm{r}\,\delta V_{\rm ind}\left(\bm{r},t\right){\rho}\left(\bm{r}\right). 
	\end{equation}
	Through standard perturbation theory, we obtain the ground state of this system as
	\begin{align}
		\left|\psi'_0\right> &= \left|\psi_0\right> + \sum_{ij}  \frac{\left|\psi_{ij}\right>\left<\psi_{ij}\right|H_I\left|\psi_0\right>}{\left(E_0-E_{ij}+\hbar\omega+i\hbar\alpha\right)},\nonumber\\
		&= \left|\psi_0\right> + \sum_{ij}  \frac{\left|\psi_{ij}\right>\left<\psi_{ij}\right|H_I\left|\psi_0\right>}{\left(\epsilon_i-\epsilon_{j}+\hbar\omega+i\hbar\alpha\right)},
	\end{align}
	where we have substituted $E_0-E_{ij} = \epsilon_i-\epsilon_{j}$.
	The change is charge density to the first order in $\delta V$ is given by
	\begin{align}
		\delta \rho \left(\bm{r},t\right) &= \left<\psi'_0\right|{\rho}\left(\bm{r}\right)\left|\psi'_0\right> - \left<\psi_0\right|{\rho}\left(\bm{r}\right)\left|\psi_0\right>, \nonumber\\
		&= \sum_{ij} \frac{\left<\psi_0\right|{\rho}\left(\bm{r}\right)\left|\psi_{ij}\right>\left<\psi_{ij}\right|H_I\left|\psi_0\right>}{\left(\epsilon_i-\epsilon_{j}+\hbar\omega+i\hbar\alpha\right)} +\sum_{ij} \frac{\left<\psi_{ij}\right|{\rho}\left(\bm{r}\right)\left|\psi_{0}\right>\left<\psi_{0}\right|H_I\left|\psi_{ij}\right>}{\left(\epsilon_i-\epsilon_{j}-\hbar\omega-i\hbar\alpha\right)}.
	\end{align}
	Substituting for $\left|\psi_0\right>$ and $\left|\psi_{ij}\right>$, and performing an ensemble average \cite{cohen_louie_2016} at finite temperature we have
	\begin{equation}
		\delta{\rho}\left(\bm{r},t\right) = \sum_{ij} f_i \left(1-f_j\right) \left[ \frac{\phi^*_i(\bm{r})\phi_j(\bm{r})\int d\bm{r}'\phi^*_j(\bm{r}')\phi_i(\bm{r}')\delta V_{\rm ind}(\bm{r}',t)}{\left(\epsilon_i-\epsilon_{j}+\hbar\omega+i\hbar\alpha\right)} + \frac{\phi_i(\bm{r})\phi^*_j(\bm{r})\int d\bm{r}'\phi_j(\bm{r}')\phi^*_i(\bm{r}')\delta V_{\rm ind}(\bm{r}',t)}{\left(\epsilon_i-\epsilon_{j}-\hbar\omega-i\hbar\alpha\right)}\right],  
	\end{equation}
	where $f_i, f_j$ are the Fermi-Dirac distribution functions. 
	From the linear response theory
	\begin{equation}
		\delta\rho\left(\bm{r},t\right) = \int d\bm{r}' \chi_0 \left(\bm{r},\bm{r}',\omega\right)\delta V_{\rm ind}\left(\bm{r}',t\right).
	\end{equation}
	Hence,
	\begin{equation}
		\chi_0 \left(\bm{r},\bm{r}',\omega\right) = \sum_{ij} f_i \left(1-f_j\right) \left[ \frac{\phi^*_i(\bm{r})\phi_j(\bm{r}) \phi^*_j(\bm{r}')\phi_i(\bm{r}')}{\left(\epsilon_i-\epsilon_{j}+\hbar\omega+i\hbar\alpha\right)} + \frac{\phi_i(\bm{r})\phi^*_j(\bm{r})\phi_j(\bm{r}')\phi^*_i(\bm{r}')}{\left(\epsilon_i-\epsilon_{j}-\hbar\omega-i\hbar\alpha\right)}\right],
	\end{equation}
	and in the reciprocal space
	\begin{align}
		\chi_0 \left(\bm{q}+\bm{G},\bm{q}'+\bm{G}',\omega\right) = \chi^{\bm{G}\bm{G}'}_0 \left(\bm{q}, \omega\right) &= \frac{1}{\Omega}\,\int d\bm{r}\, e^{-i\bm{q}\cdot\bm{r}}e^{-i\bm{G}\cdot\bm{r}}\int d\bm{r}'  \chi_0 \left(\bm{r},\bm{r}',\omega\right)   e^{i\bm{q}'\cdot\bm{r}'}e^{i\bm{G}'\cdot\bm{r}'},\nonumber\\
		\chi^{\bm{G}\bm{G}'}_0 \left(\bm{q}, \omega\right) = \frac{1}{\Omega} \sum_{n,n',\bm{k}\sigma} f_{n\bm{k}} \left(1-f_{n'\bm{k+q}}\right) \Big[& \frac{\left<n,\bm{k}\right|e^{-i\left(\bm{q}+\bm{G}\right)\cdot\bm{r}}\left|n',\bm{k}+\bm{q}\right>\left<n',\bm{k}+\bm{q}\right|e^{i\left(\bm{q}+\bm{G}'\right)\cdot\bm{r}'}\left|n,\bm{k}\right>}{\left(\epsilon_{n\bm{k}}-\epsilon_{n'\bm{k}+\bm{q}}+\hbar\omega+i\hbar\alpha\right)} \nonumber\\&+ \frac{\left<n',\bm{k}+\bm{q}\right|e^{i\left(\bm{q}+\bm{G}\right)\cdot\bm{r}}\left|n,\bm{k}\right>\left<n,\bm{k}\right|e^{-i\left(\bm{q}+\bm{G}'\right)\cdot\bm{r}'}\left|n',\bm{k}+\bm{q}\right>}{\left(\epsilon_{n\bm{k}}-\epsilon_{n'\bm{k}+\bm{q}}-\hbar\omega-i\hbar\alpha\right)}\Big],    
	\end{align}
	where, $\Omega$ is the crystal volume, \mbox{$f_{n\bm{k}} = [e^{{(\epsilon_{nk}-\epsilon_f})/k_B T} +1]^{-1}$}, $\bm{k}$ and $\sigma$ are the carrier momentum and spin, respectively. Substituting the above expression in Eq.~(\ref{eq:longitudinaldielectric}), we can obtain the longitudinal dielectric function (density-density response function) $\varepsilon^{\bm{G}\bm{G}'}_L$. This response determines the plasmon screening in a material. The off-diagonal elements $\bm{G}\neq \bm{G}' \neq 0$ represents the local-field effects \cite{Adler} on the dielectric response of a material.    
	
	Within the relaxation time approximation (RPA), we neglect the exchange correlation term $K_{\rm xc}$. Hence, the longitudinal dielectric function
	\begin{align}
		\varepsilon_L^{\bm{G}\bm{G}'}(\bm{q},\omega) &= 1 - V_c \chi_0,\\
		&= \delta_{\bm{G}\bm{G}'}-\frac{4\pi e^2}{|\bm{q}+\bm{G}|^2}\,\frac{1}{\Omega} \sum_{n,n',\bm{k}\sigma} f_{n\bm{k}} \left(1-f_{n'\bm{k+q}}\right) \Big[\frac{\left<n,\bm{k}\right|e^{-i\left(\bm{q}+\bm{G}\right)\cdot\bm{r}}\left|n',\bm{k}+\bm{q}\right>\left<n',\bm{k}+\bm{q}\right|e^{i\left(\bm{q}+\bm{G}'\right)\cdot\bm{r}'}\left|n,\bm{k}\right>}{\left(\epsilon_{n\bm{k}}-\epsilon_{n'\bm{k}+\bm{q}}+\hbar\omega+i\hbar\alpha\right)} + c.c\Big].\nonumber
	\end{align}
	For an insulating or semiconducting system, the Fermi-Dirac distribution function can be approximated by a step function. Hence, we can simplify the above expression as
	\begin{align}
		\varepsilon_L^{\bm{G}\bm{G}'}(\bm{q},\omega) 
		&= \delta_{\bm{G}\bm{G}'}-\frac{8\pi e^2}{q^2}\,\frac{1}{\Omega} \sum_{c,v,\bm{k}}  \Big[\frac{\left<c,\bm{k}\right|e^{-i\left(\bm{q}+\bm{G}\right)\cdot\bm{r}}\left|v,\bm{k}+\bm{q}\right>\left<v,\bm{k}+\bm{q}\right|e^{i\left(\bm{q}+\bm{G}'\right)\cdot\bm{r}'}\left|c,\bm{k}\right>}{\left(\epsilon_{c,\bm{k}}-\epsilon_{v,\bm{k}+\bm{q}}+\hbar\omega+i\hbar\alpha\right)} + c.c\Big],\label{eq:longitudinaldielectricexp}
	\end{align}
	where, $c$ and $v$ represents the conduction and valence band index, respectively, and a factor of $2$ in the numerator here is coming from the spin index.
	
	\section{Transverse Dielectric Function}\label{sec:transversedielectric}
	In the previous section, we derived the expression for longitudinal dielectric function corresponding to the change in the scalar potential. In this section, we derive the transverse dielectric function (current-current correlation function) that will determine the dielectric response of a material to a transverse electromagnetic probe. We would like to obtain the transverse dielectric function as a function of both $\omega$ and $\bm{q}$, including the local field effects. The electromagnetic displacement vector is 
	\begin{align}
		\bm{D}_T(\bm{r},t) = \bm{E}_T(\bm{r},t) + 4\pi \bm{P}_T(\bm{r},t),
	\end{align}
	where $\bm{P}_T$ is the polarization vector. In case of linear dielectric materials, \mbox{$\bm{D}_T(\bm{r},t) =\int d\bm{r}'\, \bm{\varepsilon}_T(\bm{r},\bm{r}',t)\cdot\bm{E}(\bm{r}',t)$}. In the frequency space, within the linear response theory one can write
	\begin{align}
		-i\omega\int d\bm{r}'\,\left(\bm{\varepsilon}_T(\bm{r},\bm{r}',\omega)-\delta(\bm{r}-\bm{r}')\bm{1}\right)\cdot\bm{E}(\bm{r}',\omega) = 4\pi\bm{J}_{\rm ind}(\bm{r},\omega),
	\end{align}
	where we have used the relation $\partial_t \bm{P}_T = \bm{J}_{\rm ind}$. We define the transverse susceptibility tensor 
	\begin{equation}
		\bm{\chi}_T(\bm{r},\bm{r}',\omega) = \bm{\varepsilon}_T(\bm{r},\bm{r}',\omega)-\delta(\bm{r}-\bm{r}')\bm{1}.
	\end{equation} 
	Without loss of generality, one can express the electric field in terms of the vector potential as, $\bm{E} = i\omega\bm{A}/c$. Hence,
	\begin{equation}\label{eq:transrealspace}
		\int d\bm{r}'\bm{\chi}_T(\bm{r},\bm{r}',\omega)\cdot\bm{A}(\bm{r}',\omega) = \frac{4\pi c}{\omega^2}\,\bm{J}_{ind}(\bm{r},\omega).
	\end{equation}
	Local-field effects arise in a material due to rapidly varying microscopic
	electric field components within a unit cell. Hence, the vector potential is taken to be of the form
	\begin{equation}
		\bm{A}(\bm{r}', \omega) = \sum_{\bm{G}',\bm{q}} A_{\bm{G}'}(\bm{q}, \omega)\,\bm{t}_{\bm{G}'}\, e^{i\left(\bm{q}+\bm{G}'\right)\cdot\bm{r}'},
	\end{equation}
	where the transverse unit vector $\bm{t}_{\bm{G}} \cdot (\bm{q}+\bm{G}) = 0$, and $\bm{q}$ is restricted within the first Brillouin zone. 
	We define in the Fourier space, the transverse susceptibility tensor
	\begin{align}\label{eq:transsusceptfourier}
		\chi_T^{\bm{G}\bm{G}'}\left(\bm{q},\omega\right) &=  \int d\bm{r}\int d\bm{r}'\,e^{-i\left(\bm{G}+\bm{q}\right)\cdot\bm{r}}\,\bm{t}_{\bm{G}}\cdot\bm{\chi}(\bm{r},\bm{r}',\omega)\cdot\bm{t}_{\bm{G}'}\, e^{i\left(\bm{q}+\bm{G}'\right)\cdot\bm{r}'}.
	\end{align}
	We would like to obtain an expression for $\chi_T^{\bm{G}\bm{G}'}(\bm{q},\omega)$ starting from the electromagnetic Hamiltonian 
	\begin{equation}
		H = \frac{\left(\bm{p}-\displaystyle\frac{e}{c}\bm{A}\right)^2}{2m} + U(\bm{r}),
	\end{equation}
	where $U(\bm{r})$ is the periodic lattice potential. The unperturbed crystal lattice satisfy the Hamiltonian \mbox{$H_0 = \bm{p}^2/2m + U(\bm{r})$} and the corresponding wavefunction is given by \mbox{$\psi_{n\bm{k}} = u_n(\bm{r})e^{i\bm{k}\cdot\bm{r}}$}. The perturbed Hamiltonian is given by 
	\begin{equation}
		H_1 = -\frac{e}{2mc}\left(\bm{p}\cdot\bm{A}+\bm{A}\cdot\bm{p}\right) + \frac{e^2}{2mc^2}\bm{A}^2.    
	\end{equation}  
	Now consider the single particle Liouville equation
	\begin{equation}
		i\hbar\frac{\partial \rho}{\partial t} = \left[H,\rho\right],
	\end{equation}
	where $\rho = \rho_0 +\rho_1$ is the single partial density matrix, and the unperturbed density matrix $\rho_0 = e\delta\left(\bm{r}-\bm{r}'\right)$. The unperturbed density matrix satisfy the eigenvalue equation of the form $\rho_0\psi_{n\bm{k}} = f_{n\bm{k}}\psi_{n\bm{k}}$, where \mbox{$f_{n\bm{k}}$} is the Fermi-Dirac distribution. Consider the expectation value 
	\begin{align}
		i\hbar\frac{\partial\left<n'\bm{k}'\right| \rho_1\left|n\bm{k}\right>}{\partial t} &= \left(\epsilon_{n'\bm{k}'}-\epsilon_{n\bm{k}}\right)\left<n'\bm{k}'\right|\rho_1\left|n\bm{k}\right>+\left(-f_{n'\bm{k}'}+f_{n\bm{k}}\right)\left<n'\bm{k}'\right|H_1\left|n\bm{k}\right>,    
	\end{align}
	and make the ansatz that the time variation of $\left<n'\bm{k}'\right| \rho_1\left|n\bm{k}\right> \sim e^{-i\omega t}$. Hence we obtain 
	\begin{align}
		\hbar\omega\left<n'\bm{k}'\right| \rho_1\left|n\bm{k}\right> &= \left(\epsilon_{n'\bm{k}'}-\epsilon_{n\bm{k}}\right)\left<n'\bm{k}'\right|\rho_1\left|n\bm{k}\right>+\left(-f_{n'\bm{k}'}+f_{n\bm{k}}\right)\left<n'\bm{k}'\right|H_1\left|n\bm{k}\right>,\nonumber\\
		\left<n'\bm{k}'\right| \rho_1\left|n\bm{k}\right>&= \frac{\left(f_{n'\bm{k}'}-f_{n\bm{k}}\right)}{\epsilon_{n'\bm{k}'}-\epsilon_{n\bm{k}}-\hbar\omega}\left<n'\bm{k}'\right|H_1\left|n\bm{k}\right>.
	\end{align}
	Expectation value of the perturbed Hamiltonian is given by
	\begin{align}
		\left<n'\bm{k}'\right|H_1\left|n\bm{k}\right> =-\frac{e}{2mc}\left<n'\bm{k}'\right|\bm{p}\cdot\bm{A}+\bm{A}\cdot\bm{p}\left|n\bm{k}\right>+\frac{e^2}{2mc^2}\left<n'\bm{k}'\right|\bm{A}^2\left|n\bm{k}\right>.
	\end{align}
	Second term is the diamagnetic term whose contribution is negligible while compared to the first paramagnetic term. 
	\begin{align}
		\left<n'\bm{k}'\right|H_1\left|n\bm{k}\right> &\simeq \frac{ie\hbar}{2mc}\int d\bm{r}\, \left[\psi^\dagger_{n'\bm{k}'}(\bm{r})\nabla\cdot\left(\bm{A}\,\psi_{n\bm{k}}(\bm{r})\right)+\psi^\dagger_{n'\bm{k}'}(\bm{r})\,\bm{A}\cdot\nabla\psi_{n\bm{k}}(\bm{r})\right],\nonumber\\
		&= \frac{ie\hbar}{2mc}\int d\bm{r}\, \left[-\nabla\psi^\dagger_{n'\bm{k}'}(\bm{r})\cdot\left(\bm{A}\,\psi_{n\bm{k}}(\bm{r})\right)+\psi^\dagger_{n'\bm{k}'}(\bm{r})\,\bm{A}\cdot\nabla\psi_{n\bm{k}}(\bm{r})\right],\nonumber\\
		&=-\frac{e}{c}\int d\bm{r}\,\bm{A}\cdot\psi^\dagger_{n'\bm{k}'}(\bm{r})\,\bm{J}_0\psi_{n\bm{k}}(\bm{r}),
	\end{align}
	where, $\bm{J}_0$ is probability current operator. 
	The transverse induced current is given by
	\begin{align}
		\bm{J}_{\rm ind}\left(\bm{r},\omega\right) &= -e{\rm Tr}\left(\bm{J}_{0}\,\rho_1\right), \nonumber\\
		&= -\frac{e}{\Omega}\sum_{n,n',\bm{k}\,\bm{k}'}\left<n\bm{k}\right|\bm{J}_{0}\left|n'\bm{k}'\right>\left<n'\bm{k}'\right|\rho_1\left|n\bm{k}\right>,\nonumber\\
		&= \frac{e^2}{\Omega c}\sum_{n,n',\bm{k}\,\bm{k}'}\psi^\dagger_{n\bm{k}}(\bm{r})\,\bm{J}_0\psi_{n'\bm{k}'}(\bm{r})\,\frac{\left(f_{n'\bm{k}'}-f_{n'\bm{k}}\right)}{\epsilon_{n'\bm{k}'}-\epsilon_{n\bm{k}}-\hbar\omega}\int d\bm{r}'\,\bm{A}\cdot\psi^\dagger_{n\bm{k}}(\bm{r}')\bm{J}_0\psi_{n'\bm{k}'}(\bm{r}').
	\end{align}
	Using Eq.~(\ref{eq:transrealspace}), the susceptibility tensor in real space is given by 
	\begin{equation}
		\chi_T\left(\bm{r},\bm{r}',t\right) =   \frac{4\pi e^2}{\Omega\, \omega^2}\sum_{n,n',\bm{k}\,\bm{k}'}\psi^\dagger_{n\bm{k}}(\bm{r})\,\bm{J}_0\psi_{n'\bm{k}'}(\bm{r})\,\frac{\left(f_{n'\bm{k}'}-f_{n\bm{k}}\right)}{\epsilon_{n'\bm{k}'}-\epsilon_{n\bm{k}}-\hbar\omega}\,\psi^\dagger_{n'\bm{k}'}(\bm{r}')\,\bm{J}_0\psi_{n\bm{k}}(\bm{r}').  
	\end{equation}
	Substituting the above relation in Eq.~(\ref{eq:transsusceptfourier}), we obtain 
	\begin{equation}
		\chi_T^{\bm{G}\bm{G}'}\left(\bm{q},\omega\right) = \frac{4\pi e^2}{\Omega\, \omega^2}\sum_{n,n',\bm{k}}\left<{n\bm{k}}\right|e^{-i\left(\bm{G}+\bm{q}\right)\cdot\bm{r}}\,\bm{t}_{\bm{G}}\cdot\bm{J}_0\left|{n'\bm{k}+\bm{q}}\right>\,\frac{\left(f_{n'\bm{k}+\bm{q}}-f_{n\bm{k}}\right)}{\epsilon_{n'\bm{k}+\bm{q}}-\epsilon_{n\bm{k}}-\hbar\omega}\,\left<{n'\bm{k}+\bm{q}}\right|\,e^{i\left(\bm{G}'+\bm{q}\right)\cdot\bm{r}'}\,\bm{t}_{\bm{G}'}\cdot\bm{J}_0\left|{n\bm{k}}\right>,    
	\end{equation}
	where we have also utilized the conservation of crystal momentum. Hence, the transverse dielectric function is given by
	\begin{align}
		\varepsilon_T^{\bm{G}\bm{G}'}(\bm{q},\omega) = \delta_{\bm{G}\bm{G}'}+ \frac{4\pi e^2}{\Omega\, \omega^2}\sum_{n,n',\bm{k}}&\left<{n\bm{k}}\right|e^{-i\left(\bm{G}+\bm{q}\right)\cdot\bm{r}}\,\bm{t}_{\bm{G}}\cdot\bm{J}_0\left|{n'\bm{k}+\bm{q}}\right>\,{\left(f_{n'\bm{k}+\bm{q}}-f_{n\bm{k}}\right)}\left<{n'\bm{k}+\bm{q}}\right|\,e^{i\left(\bm{G}'+\bm{q}\right)\cdot\bm{r}'}\,\bm{t}_{\bm{G}'}\cdot\bm{J}_0\left|{n\bm{k}}\right>\times\nonumber\\&\left[{\rm P. V.}\left(\frac{1}{{\epsilon_{n'\bm{k}+\bm{q}}-\epsilon_{n\bm{k}}-\hbar\omega}}\right)+i\pi \delta\left({\epsilon_{n'\bm{k}+\bm{q}}-\epsilon_{n\bm{k}}-\hbar\omega}\right)\right],\nonumber
	\end{align}
	The real and imaginary part of the transverse dielectric function is given by
	\begin{align}
		{\rm Re}\left[\varepsilon_T^{\bm{G}\bm{G}'}\right] = \delta_{\bm{G}\bm{G}'}+\frac{8\pi e^2 \hbar^2}{\Omega} \sum_{n,n',\bm{k}}  \frac{\left(f_{n'\bm{k}+\bm{q}}-f_{n\bm{k}}\right)}{\left(\epsilon_{n'\bm{k}+\bm{q}}-\epsilon_{n\bm{k}}\right)}\Big[\frac{\left<n,\bm{k}\right|e^{-i\left(\bm{q}+\bm{G}\right)\cdot\bm{r}}\bm{t}_{\bm{G}}\cdot\bm{J}_0\left|n',\bm{k}+\bm{q}\right>\left<n',\bm{k}+\bm{q}\right|e^{i\left(\bm{q}+\bm{G}'\right)\cdot\bm{r}'}\bm{t}_{\bm{G}'}\cdot\bm{J}_0\left|n,\bm{k}\right>}{\left(\epsilon_{n\bm{k}}-\epsilon_{n'\bm{k}+\bm{q}}\right)^2-\hbar^2\omega^2}\Big],   
	\end{align}
	\begin{align}
		{\rm Im}\left[\varepsilon_T^{\bm{G}\bm{G}'}\right] = \frac{4\pi^2 e^2}{\Omega\, \omega^2}\sum_{n,n',\bm{k}}\left<{n\bm{k}}\right|e^{-i\left(\bm{G}+\bm{q}\right)\cdot\bm{r}}\,&\bm{t}_{\bm{G}}\cdot\bm{J}_0\left|{n'\bm{k}+\bm{q}}\right>\,{\left(f_{n'\bm{k}+\bm{q}}-f_{n\bm{k}}\right)}\times\nonumber\\&\left<{n'\bm{k}+\bm{q}}\right|\,e^{i\left(\bm{G}'+\bm{q}\right)\cdot\bm{r}'}\,\bm{t}_{\bm{G}'}\cdot\bm{J}_0\left|{n\bm{k}}\right> \delta\left({\epsilon_{n'\bm{k}+\bm{q}}-\epsilon_{n\bm{k}}-\hbar\omega}\right).
	\end{align}
	These relations are obtained using the Kramers-Kronig condition
	\begin{equation}
		{\rm Im}\left[\varepsilon_T^{\bm{G}\bm{G}'}(\bm{q},\omega)\right] = \frac{2\omega}{\pi}\int_{0}^{\infty} d\omega'\ \frac{\left({\rm Re}\left[\varepsilon_T^{\bm{G}\bm{G}'}(\bm{q},\omega')\right]-{\rm Re}\left[\varepsilon_T^{\bm{G}\bm{G}'}(\bm{q},\omega)\right]\right)}{\omega'^2-\omega^2}.
	\end{equation}
	In case of semiconductors and insulators, we can simplify the expression for $\varepsilon_T^{\bm{G}\bm{G}'}$ as
	\begin{align}
		\varepsilon_T^{\bm{G}\bm{G}'}(\bm{q},\omega) = \delta_{\bm{G}\bm{G}'}+ \frac{4\pi e^2}{\Omega\, \omega^2}\sum_{c,v,\bm{k}}&\left<{c\bm{k}}\right|e^{-i\left(\bm{G}+\bm{q}\right)\cdot\bm{r}}\,\bm{t}_{\bm{G}}\cdot\bm{J}_0\left|{v\bm{k}+\bm{q}}\right>\,\left<{v\bm{k}+\bm{q}}\right|\,e^{i\left(\bm{G}'+\bm{q}\right)\cdot\bm{r}'}\,\bm{t}_{\bm{G}'}\cdot\bm{J}_0\left|{c\bm{k}}\right>\times\nonumber\\&\left[{\rm P. V.}\left(\frac{1}{{\epsilon_{v\bm{k}+\bm{q}}-\epsilon_{c\bm{k}}-\hbar\omega}}\right)+i\pi \delta\left({\epsilon_{v\bm{k}+\bm{q}}-\epsilon_{c\bm{k}}-\hbar\omega}\right)\right].
	\end{align}
	
	\section{Atomistic Dielectric Function of Silicon: Isotropic Nearly-free Electron Model}\label{sec:isotropicmodel1}
	In this section, we obtain the atomistic dielectric function of silicon based on a nearly-free electron bandstructure. 
	Silicon has the diamond cubic crystal structure and the first Brillouin zone has the shape of a truncated octahedron. It has been shown earlier \cite{WalterCohen_wave} that the wavevector dependent dielectric function in diamond-type materials is insensitive to the direction of $\bm{q}$. Hence, we can replace the truncated octahedron shape of the first Brilloin zone by a sphere and obtain the dielectric properties through an isotropic model as described in the main text. We have shown that the results obtained through an isotropic nearly-free electron bandstructure agrees well with the exact band models for silicon based on plane-wave methods \cite{waltercohen_wavefreq}.
	
	A nearly-free electron model employed here was first introduced by Penn \cite{Dpenn}. This model allows for the formation of standing waves at the Brillouin zone boundaries and accounts for the Umklapp processes. In this scheme, the eigen energy and wavefunctions of an electron is given by 
	\begin{align}
		E_{\bm{k}}^{\pm} &= \frac{1}{2}\left[E_{\bm{k}}^0+E_{\bm{k}'}^{0}\pm\sqrt{\left(E_{\bm{k}}^{0}-E_{\bm{k}'}^{0}\right)^2+E_g^2} \right],\nonumber\\
		\psi_{\bm{k}}^{\pm} &= \frac{\left(e^{i\bm{k}\cdot\bm{r}}+\alpha_{\bm{k}}^{\pm}e^{i\bm{k}'\cdot\bm{r}}\right)}{\sqrt{1+\left(\alpha_{\bm{k}}^{\pm}\right)^2}},
	\end{align}
	where, 
	\begin{align}
		\alpha_{\bm{k}}^{\pm} &= \frac{E_g}{2\left(E_{\bm{k}}^\pm - E_{\bm{k}'}^0\right)}, \nonumber\\
		E_{\bm{k}}^0 &= \frac{\hbar^2 k^2}{2m}, \nonumber\\
		\bm{k}' &= \bm{k}-\bm{G}_1,\nonumber
	\end{align}
	$\bm{G}_1 = 2k_f\hat{k}$, $k_f$ is the valence Fermi wavevector, and $E_g$ is the bandgap of the material. Superscripts $+$ and $-$ represents the $k > k_f$ (conduction) and $k < k_f$ (valence) bands, respectively.  
	
	For calculation convenience we perform the change of variables, \mbox{$y = 1-{k}/{k_f}$}, \mbox{$\eta = {q}/{k_f}$}, \mbox{$\Delta = {E_g}/{4 E_F}$}, and $z = \cos\theta$. With this transformation, we obtain
	\begin{align}\label{eq:energyandwaveunitchange}
		E_{\bm{k}}^{\pm} &= E_F \left[\left(1-y\right)^2+2y\pm 2\sqrt{y^2+\Delta^2}\right],\nonumber\\
		E_{\bm{k}+\bm{q}}^{\pm} &= E_F\left[\left(1-y\right)^2+\eta^2 +2y\left(1-\eta z\right)\pm 2\sqrt{\left(\eta z - y\right)^2+\Delta^2}\right],\nonumber\\
		\alpha_{\bm{k}}^\pm &= \frac{\Delta}{-y\pm \sqrt{y^2+\Delta^2}},\ \ \alpha_{\bm{k}+\bm{q}}^\pm = \frac{\Delta}{\eta z-y\pm \left(\sqrt{(\eta z - y)^2+\Delta^2}\right)}. 
	\end{align}
	
	We will now proceed to obtain the longitudinal and transverse dielectric function of silicon using this model. Through inspection, we see that for either case, within this model only the dielectric matrix elements corresponding to $\bm{G} = 0$ and $\bm{G}_1 = 2k_f \hat{k}$ are non-zero. All higher order elements corresponding to the reciprocal lattice vectors vanish.
	
	For the nearly-free electron bandstructure, the longitudinal dielectric function can be simplified as 
	\begin{align}
		\varepsilon_L^{mn}(\bm{q},\omega) 
		&= 1-\frac{8\pi e^2}{q^2}\,\frac{1}{\Omega}\,\frac{1}{3}\,\frac{\Omega}{(2\pi)^3}\int d^3 k\, \Bigg[\frac{\left<\bm{k}\right|e^{-i\left(\bm{q}+\bm{G}_m\right)\cdot\bm{r}}\left|\bm{k}+\bm{q}\right>\left<\bm{k}+\bm{q}\right|e^{i\left(\bm{q}+\bm{G}_n\right)\cdot\bm{r}'}\left|\bm{k}\right>}{\left(E^+_{\bm{k}}-E^-_{\bm{k}+\bm{q}}+\hbar\omega+i\hbar\alpha\right)} + c.c\Bigg],
	\end{align}
	and the transverse dielectric function is given by
	\begin{align}
		\varepsilon_T^{mn}(\bm{q},\omega) = \delta_{mn}+ \frac{4\pi e^2}{\Omega\, \omega^2}\,\frac{1}{3}\,\frac{\Omega}{(2\pi)^3}\int d^3 k\,&\left<{\bm{k}}\right|e^{-i\left(\bm{G}_m+\bm{q}\right)\cdot\bm{r}}\,\bm{t}_{\bm{G}_m}\cdot\bm{J}_0\left|{\bm{k}+\bm{q}}\right>\,\left<{\bm{k}+\bm{q}}\right|\,e^{i\left(\bm{G}_n+\bm{q}\right)\cdot\bm{r}'}\,\bm{t}_{\bm{G}_n}\cdot\bm{J}_0\left|{\bm{k}}\right>\times\nonumber\\&\left[{\rm P. V.}\left(\frac{1}{{E^-_{\bm{k}+\bm{q}}-E^+_{\bm{k}}-\hbar\omega}}\right)+i\pi \delta\left({E^-_{\bm{k}+\bm{q}}-E^+_{\bm{k}}-\hbar\omega}\right)\right],
	\end{align}
	where we have replaced $\sum_{\bm{k}}\rightarrow \Omega/(2\pi)^3\int d^3k$, $1/3$ factor is introduced due to isotropic model, and indices \mbox{$m,n = 0, 1$}. Hence, within this model, we obtain both longitudinal and transverse dielectric function in a $2\times 2$ matrix form. Band parameters employed in our calculation are tabulated in Table~\ref{tab:bandparams}. Integrating over the Brillouin zone are performed numerically after substituting for the energy and wavefunction in Eq.~(\ref{eq:energyandwaveunitchange}). 
	
\begin{figure}
		\centering
		\includegraphics[width = 3in]{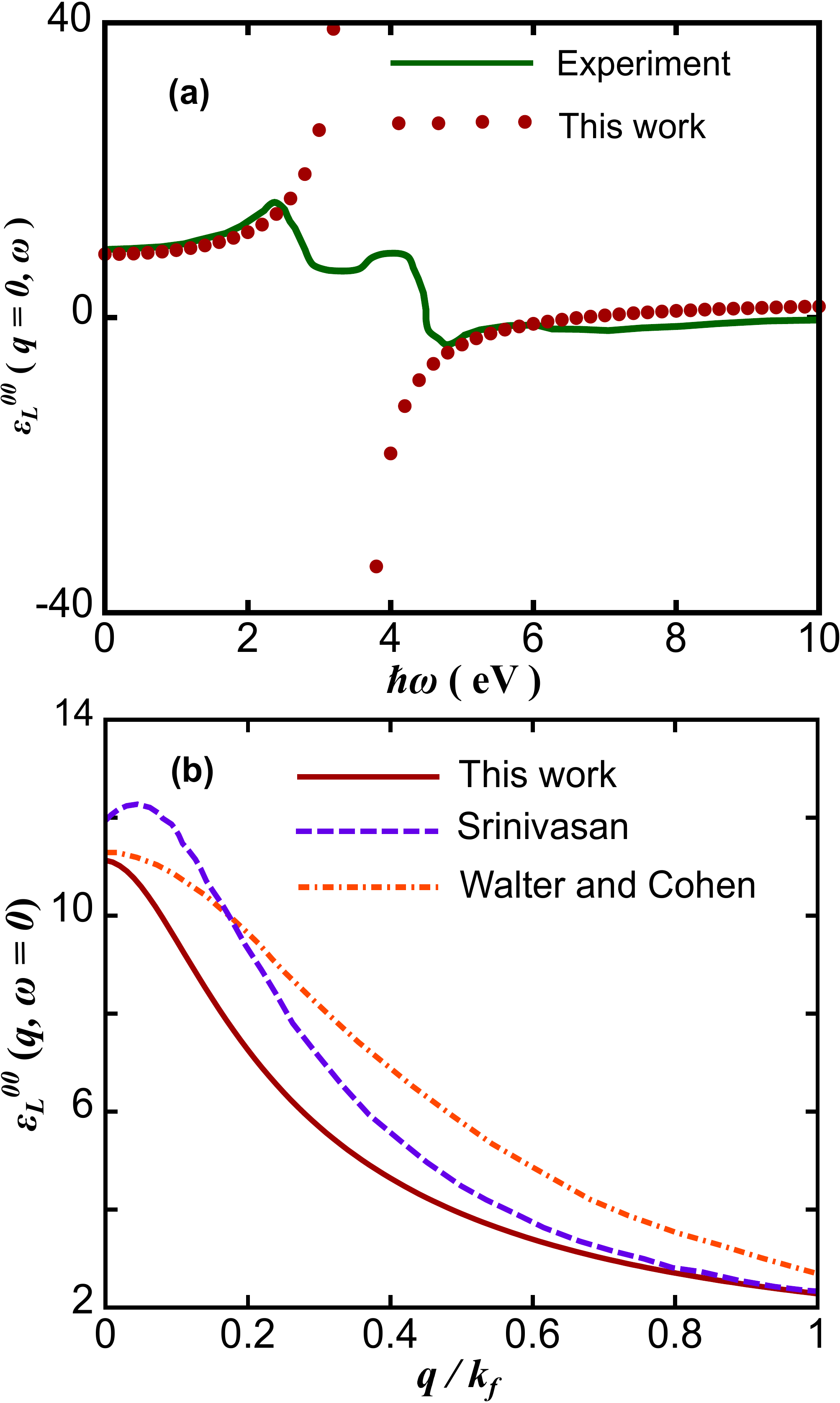}
		\caption{(a) Dielectric function obtained through the isotropic model is compared with the experimental data \cite{PhilippEhrenreich} at zero momentum. (b) Dielectric function as a function of momentum is plotted at zero frequency and compared with the work by Srinivasan \cite{Srinivasan}, and Walter {\it et al.}, \cite{WalterCohen_wave}. Isotropic model used in our calculations has an excellent match with the experimental data and the full bandstructure analysis.}
		\label{fig:longdielectric_compare}
	\end{figure}

	Our calculations for $\varepsilon^{00}_L$ matches well with the earlier calculations by Srinivasan \cite{Srinivasan}, and by Walter and Cohen \cite{waltercohen_wavefreq} (Fig.~\ref{fig:longdielectric_compare}(b)). Across the frequency range, calculations through this isotropic nearly-free electron model has excellent match with the experimentally measured dielectric function as well (Fig.~\ref{fig:longdielectric_compare}(a)). This calculation can be extended to calculate the dielectric functions for all frequencies and wavevectors as shown in the main text. Here, $\varepsilon_{L,T}^{ij}(q \neq 0, \omega)$ represents the non-local contributions to the dielectric properties. $\varepsilon_{L,T}^{01}, \varepsilon_{L,T}^{11}$ are due to the local-field effects. In literature, typically only $\varepsilon^{00}_L$ is calculated and used to obtain all dielectric properties of the materials. Our calculations show that the higher-order dielectric components have significant contributions even at zero frequency.   
	
	\begin{table}[h]
		\vspace{0.1in}
		\centering
		\begin{tabular}{c c c c c}
			\hline\\
			Si     &  $E_g$ & $E_F$ & $k_f$ & $\Delta$ \\[3pt]
			\hline\\
			& 3.84\,eV & 12.0\,eV & 1.78\,$\angstrom^{-1}$ & 0.07036\\[6pt]
			\hline
		\end{tabular}
		\caption{Band parameters employed in our calculations to obtain the dielectric properties of Silicon \cite{Auluck} are tabulated.}
		\label{tab:bandparams}
	\end{table}
	
	\section{Maxwell Hamiltonian in Silicon: Isotropic Nearly-Free Electron Model}
	For an isotropic electron bandstructure, $\varepsilon_T^{\bm{G}\bm{G}'}$ reduces to a $2\times 2$ matrix as shown in Sec.~\ref{sec:isotropicmodel}. Within this scheme, the Maxwell hamiltonian equation of motion reduces to a simpler form
	\begin{equation}\label{eq:coupledEMatomistic}
		\left|\bm{q}+\bm{G}_i\right|^2 \bm{E}_{i} = \frac{\omega^2}{c^2}\sum_{j = 0, 1} \varepsilon_T^{ij}(\bm{q},\omega) \bm{E}_{j},
	\end{equation}
	where, $\bm{G}_0 = 0$, $\bm{G}_1 = 2k_f\hat{k}$. 
	Solutions to the above equation results in the anomalous atomistic electrodynamic dispersion discussed in the main text. 
	The corresponding electric field solutions will have the Bloch expansion form
	\begin{align}\label{eq:efiledatomistic}
		\bm{E}(\bm{r},\omega) &= e^{i\bm{q}\cdot\bm{r}} \left[E_0 + E_1\, e^{i{2k_f \hat{k}}\cdot\bm{r}} \right]\hat{q}_{\perp},\nonumber
		\\&= e^{i\bm{q}\cdot\bm{r}}{E_N}\left[1+e^{i{2k_f \hat{k}}\cdot\bm{r}}\,\beta(\bm{q},\omega)\right]\hat{q}_{\perp},
	\end{align}
	where, $E_N$ is the amplitude, and $\beta(\bm{q},\omega)$ is the atomic modulation function given by
	\begin{equation}
		\beta(\bm{q},\omega) = \frac{\left[q^2 \displaystyle\frac{c^2}{\omega^2(\bm{q})}-\varepsilon^T_{00}(\bm{q}, \omega)\right]}{\varepsilon^T_{01}(\bm{q}, \omega)}.
	\end{equation}
	Atomistic modulation function is nearly zero for the regular band, whereas vary significantly in the anomalous band. Hence, the anomalous band has significant contributions from the higher order reciprocal lattice components. 
	
	\section{Transmission Coefficient}
	In this section, we derive the expression for total transmission coefficient including both regular and anomalous band contributions. We consider an electromangetic wave from vacuum injected at a normal angle on a silicon block of very large thickness. Inside the material, for energies $\hbar \omega < 4.3\,$eV, transmitted wave has contribution only from the regular band. However, for energies $\hbar \omega > 4.3\,$eV, the total transmitted wave has contributions from both regular and anomalous band. 
	
	Let the region $y > 0$ to be vacuum and $y < 0$ is occupied by silicon. First, let us consider the case $\hbar \omega < 4.3\,$eV. The incident and reflected plane waves are given by
	\begin{align}
		\bm{E}_1(y, t) &=  E_I\,e^{-iq_0y}e^{-i\omega t}\,\hat{z}, \nonumber\\
		\bm{E}_{2}(y, t) &= E_{2}\,e^{ iq_0y}e^{-i\omega t}\,\hat{z},\label{eq:efieldincident}
	\end{align}
	where, $q_0$ is the free field wavevector that satisfy the relation, \mbox{$q_0^2 = \varepsilon_0\, \omega^2/c^2$}. Atomistic modulation function is nearly zero in regular band and hence the transmitted field is given by
	\begin{equation}\label{eq:efieldtransmitted}
		\bm{E}_{3,R}(y, t) = E_{3,R}\,e^{-iq_R y} e^{-i\omega t}\,\hat{z},
	\end{equation}
	where, $q_R$ is the wavevector of the regular band at a given frequency $\omega$ derived from the Maxwell Hamiltonian. The tangential component of $\bm{E}$ and $\bm{H}$ are continuous across the interface $y = 0$. Hence
	\begin{align}
		E_1 + E_{2} &= E_{3,R},\nonumber\\
		{q_0}\,\left(E_1-E_{2}\right) &= {q_R}\,E_{3,R}.
	\end{align} 
	Therefore the amplitude of transmitted wave is given by
	\begin{equation}\label{eq:regularamplitude}
		E_{3,R} = E_1 \frac{2}{\left(1+\displaystyle\frac{q_R}{q_0}\right)}. 
	\end{equation}
	The total transmission coefficient is defined as
	\begin{equation}
		T_{\rm total} = \frac{\left<\bm{S}_{3,R}\cdot \hat{z}\right>\Big{|}_{y=0}}{\left<\bm{S}_{1}\cdot \hat{z}\right>\Big{|}_{y=0}},
	\end{equation}
	where, the pointing vector $\bm{S}_{3,R} = (1/2)\,{\rm Re}\left[\bm{E}_{3,R}\times \bm{H}_{3,R}^\dagger\right]$ and $\bm{S}_{1} = (1/2)\,{\rm Re}\left[\bm{E}_{1}\times \bm{H}^\dagger_{1}\right]$. Substituting Eq.~(\ref{eq:regularamplitude}) we obtain
	\begin{align}
		{T}_{\rm total} = \frac{q}{q_0} \left|\frac{E_{3,R}}{E_1}\right|^2 = \frac{q}{q_0}\frac{4}{\left(1+\displaystyle\frac{q}{q_0}\right)^2}.
	\end{align}
	In the macroscopic limit, $q \approx \sqrt{\varepsilon^{00}_T(\omega)}\,\omega/c$. In this limit, ${T}_{\rm total}$ reduces to the standard form
	\begin{equation}
		{T}_{\rm total} \approx \left[\frac{4\sqrt{\varepsilon^{00}_T(\omega)}}{(\sqrt{\varepsilon^{00}_T(\omega)}+1)^2}\right].    
	\end{equation}
	
	Next, we consider the case of $\hbar\omega > 4.3\,$eV. Above this energy, both regular and anomalous band contributes to the total transmission spectrum. Hence, the total transmitted field has the form
	\begin{align}
		\bm{E}_{3} &= E_{3,R} e^{-iq_R y} e^{-i\omega t}\,\hat{z}+{E}_{3,A}\,e^{-iq_A y}\left(1+\beta(q_A, \omega)e^{-i2k_f y}\right) e^{-i\omega t}\,\hat{z},\nonumber\\
	\end{align}
	where, $q_A$ is the wavevector at a given frequency $\omega$ corresponding to the anomalous band derived from the Maxwell Hamiltonian, $E_{3,A}$ is the amplitude of the anomalous band. 
	The continuity conditions at the interface $y = 0$ including the anomalous contributions are given by
	\begin{align}
		E_1 + E_{2}  &= E_{3,A}(1+\beta(q_A,\omega)),\label{eq:continuityar1}\\
		{q_0}\left(E_1-E_{2}\right) &= {q_R}\,E_{3,R} + E_{3,A}{\left(q_A+(q_A+2k_f)\beta(q_A, \omega)\right)}.\label{eq:continuityar2}   
	\end{align}
	Along the above two relations, energy conservation requires that the incident electromagnetic intensity $I_{1}$ is equal to the sum of reflected $(I_2)$ and transmitted $(I_3)$ intensity at the interface $y=0$. 
	\begin{equation}\label{eq:intensity}
		I_1(y=0) = I_2(y=0) + I_3(y=0),     
	\end{equation}
	where,
	\begin{align}
		I_1 &=\left<\bm{S}_{1}\cdot \hat{z}\right> = \frac{q_0}{2}\,E^2_1, \nonumber\\
		I_2 &=\left<\bm{S}_{2}\cdot \hat{z}\right> = \frac{q_0}{2}\,E^2_2, \nonumber\\
	\end{align}
	and 
	\begin{align}
		I_3 &= \left<\bm{S}_{3}\cdot \hat{z}\right>\nonumber\\
		=& \frac{1}{2}\Bigg[q_R E_{3,R}^2 + q_A E_{3,A}^2 + (q_A + 2k_f)\beta^2(q_A, \omega) E_{3,A}^2 + 2 \cos\left[2k_f y\right]\left(q_A + k_f\right)\beta(q_A, \omega)E^2_{3,A}\nonumber\\&\hspace{0.1in}+  \cos\left[(q_R-q_A)y\right](q_R + q_A)E_{3,R}E_{3,A}\nonumber\\&\hspace{0.1in}+ \cos\left[(q_R-q_A-2k_f)y\right](q_R + q_A + 2k_f)\beta(q_A, \omega)E_{3,R}E_{3,A}\Bigg].
	\end{align}
	Solving for Eqs.~(\ref{eq:continuityar1}), (\ref{eq:continuityar2}), and (\ref{eq:intensity}) we obtain the total transmission coefficient 
	\begin{equation}
		T_{\rm total} = \frac{\left<\bm{S}_{T}\cdot \hat{z}\right>\Big{|}_{y=0}}{\left<\bm{S}_{1}\cdot \hat{z}\right>\Big{|}_{y=0}}. 
	\end{equation}
	This total transmission coefficient is derived from the atomistic nonlocal electrodynamic theory and shows a clear difference while compared to the macroscopic theory in deep ultraviolet regime, as discussed in the main text. 
	
\end{document}